\documentclass{article}
\usepackage{romp,amssymb,amsfonts,amsmath,diagrams,dsfont,amsthm}

\newtheorem{corollary}{Corollary}

\newenvironment{remark}[1][Remark]{\begin{trivlist}
\item[\hskip \labelsep {\bfseries #1}]}{\end{trivlist}}
\newenvironment{warning}[1][Warning]{\begin{trivlist}
\item[\hskip \labelsep {\bfseries #1}]}{\end{trivlist}}

\newcommand{\Dleft}{[\hspace{-1.5pt}[}
\newcommand{\Dright}{]\hspace{-1.5pt}]}
\newcommand{\SN}[1]{\Dleft #1 \Dright}

\DeclareMathOperator{\End}{End}
\DeclareMathOperator{\Vect}{Vect}

\DeclareMathOperator{\w}{w}

\title{ Tulczyjew triples and higher Poisson/Schouten structures on Lie algebroids }
\author{ Andrew James Bruce \\ School of Mathematics, The University of Manchester, Manchester, UK\\
M13 9PL  \\ e-mail: andrewjames.bruce@physics.org \\[2ex]
                      }

\begin{document}
\bibliographystyle{plain}

\maketitle
\begin{abstract}
 We show how to extend  the construction of Tulczyjew triples to   Lie algebroids via graded manifolds. We also provide a generalisation of triangular Lie bialgebroids as higher Poisson and Schouten structures \emph{on} Lie algebroids.
\end{abstract}


\section{Introduction}\label{section1}
\noindent Tulczyjew \cite{Tulczyjew:1974} showed that the natural identification $\alpha: T(T^{*}N) \rightarrow T^{*}(TN)$ should be considered as a symplectomorphism. More specifically, one can lift the canonical symplectic structure on $T^{*}N$ to $T(T^{*}N)$ and then $\alpha$ becomes a symplectomorphism from this lifted structure to the canonical symplectic structure on $T^{*}(TN)$. This is best understood using the language of double vector bundles.  It is well-known that  the double vector bundle morphism $R: T^{*}(T^{*}N) \rightarrow T^{*}(TN)$ decomposes as $T^{*}(T^{*}N) \rightarrow T(T^{*}N) \rightarrow T^{*}(TN)$. This triple of double vector bundles will be referred to  as the \emph{classical Tulczyjew triple}.\\

\noindent In this paper we construct analogues of the Tulczyjew  triple for Lie algebroids in terms of graded manifolds. Recall that a  graded manifold  is defined as a supermanifold equipped  with a privileged class of atlases where the coordinates are assigned weights taking values in $\mathds{Z}$ and the coordinate transformations are polynomial in coordinates with nonzero weights respecting the total weight, see Voronov \cite{Voronov:2001qf} for further particulars. Generally, the weight is  independent of the Grassmann parity.  The weight then extends to geometric objects over the supermanifold.    \\

\noindent  We  use these constructions to make an initial  study of higher Poisson and higher Schouten\footnote{Schouten  structures are also known as odd Poisson or Gerstenhaber structures. We will stick to the nomenclature Schouten. They are the Grassmann odd analogue of Poisson structures. } structures on Lie algebroids, generalising some of the constructions of Khudaverian \& Voronov \cite{khudaverdian-2008,voronov-2004,voronov-2005}. Such structures are understood as  higher order generalisations of the second order ``classical structures". For example, a higher Poisson structure can be thought of as the replacement of a Poisson bi-vector with an even parity inhomogeneous multivector field. In passing to the higher structures one leaves the world of Lie algebras and enters the world of their homotopy relatives, the $L_{\infty}$-algebras.   This should also be compared with the  generalized Poisson structures of de  Azc$\acute{\textnormal{a}}$rraga et.al. \cite{deAzcarraga:1996zk,deAzcarraga:1996jk}.\\

\noindent In accordance with standard practice in ``super-mathematics", we will generically drop the prefix  \emph{super}. For example, by manifold we will explicitly mean (smooth) supermanifold. All objects will be $\mathds{Z}_{2}$-graded. We will denote the parity of an object as \emph{tilde}; $\widetilde{A} \in \mathds{Z}_{2}$. By \emph{even} and \emph{odd} we will be referring to  parity and not weight. For a given vector bundle $E \rightarrow M$ the reverse parity functor $\Pi$  produces another vector bundle $\Pi E \rightarrow M$ with the fibre coordinates shifted in Grassmann parity in relation to the original vector bundle,  the weight remains unchanged. We denote the weight of an object $A$ by $\w(A)\in \mathds{Z} $. We will use the notation $E[-1]\rightarrow M$ to denote the vector bundle whose fibre coordinates have been shifted in weight by minus one in relation to the weight of the fibre coordinates on the original vector bundle.  \\

\noindent  Lie algebroids can be understood   in terms of a Lie bracket on the space of sections of a vector bundle $E \rightarrow M$ and an anchor, which is a Lie algebra morphism to vector fields $\mathfrak{X}(M)$.   Alternatively, a Lie algebroid structure  is equivalent to  a homological vector field on $\Pi E$, the ``de Rham differential". For our purposes we will take the view point that a Lie algebroid structure is equivalent to

\begin{enumerate}{}
\item A weight minus one Schouten structure on the manifold $\Pi E^{*}$
\item A weight minus one Poisson structure on the manifold $E^{*}$
\end{enumerate}

\noindent The weighs are relative to the natural weight on $E^{*}\rightarrow M$. It is the existence of these structures that allows us to define all the necessary maps to construct the graded analogues of the Tulczyjew triple for Lie algebroids.  Furthermore, these structures are required to define higher Poisson and higher Schouten structures on the Lie algebroid. Thus, much of the theory developed here will not pass over to more general vector bundles.  The details of this will be presented in this paper. \\

\noindent This paper is arranged as follows:\\

\noindent In Section \ref{prelinimaries} we recall the basic theory of double vector bundles, Lie algebroids and $L_{\infty}$-algebras as needed in later sections. In particular we will recall the definitions of homotopy Poisson, Schouten and BV- algebras as used throughout this work. There are no new results in this section.\\

\noindent In Section \ref{tulczyjew triples} we proceed to define graded analogues of the Tulczyjew triple for Lie algebroids. It should be noted that  analogues of Tulczyjew triple   for Lie algebroids have already been discussed in the literature: Mackenzie \& Xu \cite{Mackenzie:1994},  Grabowski \& Urba$\textnormal{\'n}$ski \cite{Gribowski:1997,Gribowski:1999} (who  discuss the slightly more general notion of what they call algebroids.) and Mackenzie \cite{mackenzie-2002}.    However, the approach taken here is inherently ``super" and makes explicit use of the Schouten and Poisson structures as functions on  (graded) supermanifolds. Precursors to this ``graded super" approach  include Voronov \cite{Voronov:2001qf,voronov-2006} and  Roytenberg \cite{roytenberg-1999}.\\

\noindent We consider the following commutative diagrams to be the Lie algebroid analogues of the Tulczyjew triple;\\

\begin{tabular}{l c  l }
\begin{diagram}[htriangleheight=30pt ]
T^{*}(\Pi E^{*})& \rTo & \Pi T(\Pi E^{*})[-1]\\
\dTo & \ruTo & \\
T^{*}(\Pi E) & &
\end{diagram}
&\hspace{30pt}  &
 \begin{diagram}[htriangleheight=30pt ]
\Pi T^{*}( E^{*})& \rTo & \Pi T( E^{*})[-1]\\
\dTo & \ruTo & \\
\Pi T^{*}(\Pi E) & &
\end{diagram}
\end{tabular}\\

\noindent All the maps in the above diagrams are double vector bundle morphisms in the category of graded manifolds.\\

\noindent In Section \ref{higher structures} we use the constructions of the previous sections to define and study higher Poisson and higher Schouten structures on Lie algebroids. A higher Poisson structure on a Lie algebroid is defined as an even function $\mathcal{P} \in C^{\infty}(\Pi E^{*})$ such that $\SN{\mathcal{P}, \mathcal{P}}_{S}=0$, where the bracket here is the Schouten structure that describes the Lie algebroid. Similarly, a higher Schouten structure is defined as an odd function $\mathcal{S} \in C^{\infty}(E^{*})$ such that $\{ \mathcal{S}, \mathcal{S} \}_{P}=0$, where the bracket is the Poisson structure that describes the Lie algebroid. \\

\noindent We provide a theorem that states that if $E\rightarrow M$ is a Lie algebroid with a higher Poisson  structure then $E^{*}\rightarrow M$ is canonically an $L_{\infty}$-algebroid. Similarly, we prove that if the Lie algebroid comes equipped with a higher Schouten structure, then $\Pi E^{*} \rightarrow M$ is an $L_{\infty}$-algebroid. In this sense, higher Poisson/Schouten structures on Lie algebroids provide a generalisation of the notion of a triangular Lie bialgebroid. \\

 \noindent We show how to associate  homotopy Poisson and Schouten algebras on  $C^{\infty}(\Pi E)$ with these structures, that is on the Lie algebroid analogue of differential  forms.   That is, we have an $L_{\infty}$-algebra in the sense of Lada \& Stasheff  \cite{Lada:1992wc} (suitably ``superised") on the algebra $C^{\infty}(\Pi E)$  such that the brackets are multi-derivations over the product.   This generalises the Koszul--Schouten bracket \cite{Koszul;1985} to the homotopy and Lie algebroid cases. \\

 \noindent An homotopy BV-algebra  on Lie algebroid forms can be constructed using the Lie derivative along a higher Poisson structure. This mimics Koszul's original construction  \cite{Koszul;1985}. We show that this series of higher Koszul--Schouten brackets is not completely independent of the earlier construction of higher Schouten brackets on Lie algebroid forms. In particular the higher Schouten brackets are understood as a ``formal classical limit" of the higher Koszul--Schouten brackets.  \\

\noindent We end this paper in Section \ref{discussion} with a short discussion. \\

\noindent \textbf{Nomenclature}

\noindent A \emph{differential form} over a manifold $M$ is understood as a function on the total space of $\Pi TM$. In general  differential forms need not be polynomial in the fibre coordinates,  commonly such forms are known as  \emph{pseudoforms}.  A  \emph{Lie algebroid form} is a function on the total space of $\Pi E$, where  $E \rightarrow M$ is a Lie algebroid. This definition is sufficient for our purposes as we will not be delving into the theory of integration.  Similarly, a multivector field over a manifold $M$ is understood as a function on the total space of $\Pi T^{*}M$. A \emph{Lie algebroid multivector} is  a function on the total space of $\Pi E^{*}$.  \\

\noindent A \emph{Poisson} $(\varepsilon = 0)$  or \emph{Schouten} $(\varepsilon = 1)$ \emph{algebra} is understood as a vector space $A$ with a bilinear associative multiplication and a bilinear operation $\{ ,\}: A \otimes A \rightarrow A$ such that:
\begin{list}{}
\item \textbf{Grading} $\widetilde{\{a,b \}_{\varepsilon}} = \widetilde{a} + \widetilde{b} + \varepsilon$
\item \textbf{Skewsymmetry} $\{a,b\}_{\varepsilon} = -(-1)^{(\tilde{a}+ \varepsilon)(\tilde{b}+ \varepsilon)} \{b,a \}_{\varepsilon}$
\item \textbf{Leibnitz Rule} $\{a,bc \}_{\varepsilon} = \{a,b \}_{\varepsilon}c + (-1)^{(\tilde{a} + \varepsilon)\tilde{b}} b \{a,c \}_{\varepsilon}$
\item \textbf{Jacobi Identity} $\sum_{cyclic\: a,b,c} (-1)^{(\tilde{a}+ \varepsilon)(\tilde{c}+ \varepsilon)}\{a,\{b,c\}_{\varepsilon}  \}_{\varepsilon}= 0$
\end{list} \vspace{10pt}
\noindent for all homogenous elements $a,b,c \in A$.\\

 \noindent A manifold $M$ such that $C^{\infty}(M)$ is a Poisson/Schouten algebra is known as a \emph{Poisson/Schouten manifold}. As the Poisson/Schouten brackets are biderivations over the functions they are  specified by contravariant tensor fields of rank two. A \emph{Poisson structure} on a manifold $M$ is understood as a bi-vector field $P \in C^{\infty}(\Pi T^{*}M)$ (quadratic in fibre coordinates), such that $\SN{P,P} = 0$. Here the brackets are the canonical Schouten brackets on $\Pi T^{*}M$ also known a the Schouten--Nijenhuis bracket.  The associated Poisson bracket is given by $\{f,g \}_{P} = (-1)^{\widetilde{f}+1}\SN{\SN{P,f},g}$,  with $f,g \in C^{\infty}(M)$. Similarly, a \emph{Schouten structure} on a manifold $M$ is an odd symmetric tensor field $S \in C^{\infty}(T^{*}M)$ quadratic in the fibre coordinates such that $\{S,S\}=0$. The associated Schouten bracket us given by $\SN{f,g}_{S}=(-1)^{\widetilde{f}+1} \{ \{ S,f  \},g  \}$, with  $f,g \in C^{\infty}(M)$. Note that non-trivial Schouten structures cannot exist on pure even manifolds.  The Jacobi identities on the brackets are equivalent to the self-commutating conditions of the structures.  \\

\noindent A \emph{Q-manifold} is understood as a (possibly graded) manifold $M$, equipped with an odd vector field  $Q \in \Vect(M)$ that \emph{squares to zero}, i.e. $Q^{2}= \frac{1}{2}[Q,Q]=0$. The vector field is referred to as the \emph{homological vector field}.\\

\noindent A \emph{gauge system} is understood as a the triple $(\mathcal{M}, \{\bullet, \bullet \}_{\varepsilon}, \mathcal{Q} )$ with $(\mathcal{M}, \{ \bullet, \bullet\}_{\varepsilon})$ being a Poisson or Schouten  manifold and $(\mathcal{M}, Q = \{\mathcal{Q}, \bullet \}_{\varepsilon})$  being a $Q$-manifold with the function (\emph{homological potential}) $\mathcal{Q} \in C^{\infty}(\mathcal{M})$  being of parity $\widetilde{\mathcal{Q}} = \varepsilon + 1$. Morphisms in the category of classical gauge systems are smooth maps  $\phi :\mathcal{M}_{1}\rightarrow \mathcal{M}_{2}$ such that $\phi^{*}\{f,g \}_{\mathcal{M}_{2}} = \{\phi^{*}f, \phi^{*}g  \}_{\mathcal{M}_{1}}$ and $\phi^{*}\mathcal{Q}_{\mathcal{M}_{2}} = \mathcal{Q}_{\mathcal{M}_{1}}$.\\

\section{Preliminaries} \label{prelinimaries}
\noindent For the benefit of the reader and to set notation let us review some basic facts about double vector bundles, Lie algebroids and $L_{\infty}$-algebras as needed in later sections. Double vector bundles and Lie algebroids were first introduced by Pradines \cite{Pradines1967}. $L_{\infty}$-algebras were introduced by Lada \& Stasheff \cite{Lada:1992wc} and then further discussed by Lada \& Markl \cite{Lada:1994mn}.  \\

\subsection{Double vector bundles}
\noindent   Recall that a  double vector bundle over base $M$ is a fibre bundle $D \rightarrow M$ with an extra structure. The local model is $U \times V_{1} \times V_{2} \times V_{12}$, where $U \subset M$ is an open subset and $V_{1}, V_{2}$ and $V_{12}$ are vector spaces. If we pick coordinates $x^{A}$ on $U$, $u^{\alpha}$ on $V_{1}$, $w^{i}$ on $V_{2}$ and $z^{a}$ on $V_{12}$ then admissible coordinate transformations are of the form;\\

\begin{equation}\label{admisiable transformations}
\nonumber \overline{x}^{A}  =  \overline{x}^{A}(x), \hspace{10pt} \overline{u}^{\alpha}  =   u^{\beta}T_{\beta}^{\:\: \alpha}(x), \hspace{10pt} \overline{w}^{i} = w^{j}T_{j}^{\:\: i}(x),
\end{equation}
\begin{equation}
\overline{z}^{a} = z^{b}T_{b}^{\:\: a}(x) + w^{i}u^{\beta}T_{\beta \:i}^{\:\:\:\:a}(x).
\end{equation}

\noindent These transformation laws imply that the diagram below exists with each edge being a vector bundle.

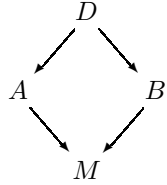
\begin{diagram}[htriangleheight=15pt ]
  &D& \\
\ldTo(1,2) & &  \rdTo(1,2)  \\
 A &  & B\\
&\rdTo(1,2)  \ldTo(1,2)&\\
&M&
\end{diagram}

 \noindent More specifically, we have $V_{1}$ as the typical fibre of the vector bundle $A \rightarrow M$, $V_{2}$ for $B \rightarrow M$, $V_{1} \times V_{12}$ for $D \rightarrow A$ and $V_{2}\times V_{12}$ for $D \rightarrow B$. The is also the core of $D$, which is the vector bundle $K \rightarrow M$ with typical fibre  $V_{12}$.\\

\noindent If we consider $D$ as a vector bundle  over $A$ we have the notion of the dual $D^{*}_{A}$, which is again a double vector bundle, see the diagram below.

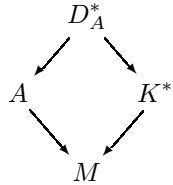
\begin{diagram}[htriangleheight=15pt ]
  &D^{*}_{A}& \\
\ldTo(1,2) & &  \rdTo(1,2)  \\
 A &  & K^{*}\\
&\rdTo(1,2)  \ldTo(1,2)&\\
&M&
\end{diagram}

\noindent If we describe $D^{*}_{A}$ using the natural local coordinates $\{x^{A}, u^{\alpha}, w_{i}, z_{a}\}$ then using the invariance of $w^{i}w_{i} + z^{a}z_{a}$ coupled with the transformation rules for $D$  the admissible coordinated changes can be written as;

\begin{equation}
 w_{i} = T_{i}^{\:\: j}(x)\overline{w}_{j} + u^{\alpha}T_{\alpha\: i}^{\:\:\:\: a}(x)\overline{z}_{a}, \hspace{15pt}
           z_{a} = T_{a}^{\:\: b}(x)\overline{z}_{b}.
\end{equation}

\noindent The ``inclusion" of the parity reversion functor $\Pi$ was laid down by Voronov in his discussion of double Lie algebroids \cite{voronov-2006}. Let $E\rightarrow M$ be a vector bundle. An important result of Voronov's study is that the canonical double vector bundle isomorphism $R: \Pi T^{*}E \rightarrow \Pi T^{*}(\Pi E)$   is an odd symplectomorphism between the canonical odd symplectic structures. \\

\subsection{Lie algebroids}
\noindent Recall the definition of a Lie algebroid as the triple $(E, [\bullet,\bullet], a )$. Here $E$ is a vector bundle over the manifold $M$ equipped with a Lie bracket acting on the module of sections $\Gamma(E)$,  together with a vector bundle morphism called the anchor $a: E \rightarrow TM$.  The anchor and the Lie bracket satisfy the following

\begin{equation}
[u,fv] = a(u)f v + (-1)^{\widetilde{f}\widetilde{u}} f[u,v],\hspace{15pt}
a([u,v]) = [a(u), a(v)],
\end{equation}

\noindent for all $u,v \in \Gamma(E)$ and $f \in C^{\infty}(M)$. To paraphrase this definition, a Lie algebroid is a vector bundle with the structure of a Lie algebra on the module of sections that can be represented by vector fields.    \\

\noindent For basis sections of $E$ the anchor and the Lie bracket are of the form

\begin{equation}
a(s_{\alpha}) = Q_{\alpha}^{A} \frac{\partial}{\partial x^{A}}, \hspace{15pt}
[s_{\alpha}, s_{\beta}] = (-1)^{\widetilde{\beta}} Q_{\alpha \beta}^{\gamma} s_{\gamma}.
\end{equation}

\noindent They satisfy the Lie algebroid structure equations;

\begin{subequations}
\begin{eqnarray}
 Q_{\alpha}^{A} \frac{\partial Q_{\beta}^{B}}{\partial x^{A}} + (-1)^{\widetilde{\alpha}\widetilde{\beta}} Q_{\beta}^{A} \frac{\partial Q_{\alpha}^{B}}{\partial x^{A}} &=& (-1)^{\widetilde{\beta}} Q^{\gamma}_{\alpha \beta}Q_{\gamma}^{B},\\
\sum _{\stackrel{\textnormal{Cyclic}}{(\alpha, \beta, \gamma)}}\left( Q_{\alpha}^{A} \frac{\partial Q_{\beta \gamma}^{\kappa}}{\partial x^{A}} + (-1)^{\beta} Q_{\alpha \beta}^{\rho} Q_{\rho \gamma}^{\kappa} \right) &=&0.
\end{eqnarray}
\end{subequations}

\noindent  It is well-known that a Lie algebroid structure on $E \rightarrow M$ is in one-to-one correspondence with a homological vector field of weight $1$ on the total space of $\Pi E$. In natural local coordinates $\{x^{A}, \xi^{\alpha} \}$ ($\widetilde{x}^{A}= \widetilde{A}, \w(x^{A}) = 0, \widetilde{\xi}^{\alpha}= \widetilde{\alpha}+1, \w(\xi^{\alpha})=1.$) the homological vector field is of the form \cite{Vaintrob:1997}

\begin{equation}
Q = \xi^{\alpha}Q_{\alpha}^{A}(x) \frac{\partial}{\partial x^{A}} + \frac{1}{2}\xi^{\alpha}\xi^{\beta}Q_{\beta \alpha}^{\gamma}(x) \frac{\partial}{\partial \xi^{\gamma}} \in \Vect(\Pi E).
\end{equation}

\noindent The Lie algebroid structure equations are directly equivalent to the homological property. If one relaxes the weight condition and considers more general homological vector fields, then we have a so-called $L_{\infty}$-algebroid. \\

\subsection{$L_{\infty}$-algebras and higher derived brackets}
\noindent  We closely follow Voronov \cite{voronov-2004} in conventions concerning $L_{\infty}$-algebras. A  vector space $V = V_{0}\oplus V_{1}$ endowed with a sequence of odd n-linear operators (which we denote as brackets) is said to be a $L_{\infty}$-algebra (c.f. \cite{Lada:1994mn,Lada:1992wc}) if
\begin{enumerate}
\item The operators are symmetric
\begin{equation}
(a_{1}, a_{2}, \cdots, a_{i},a_{j}, \cdots , a_{n}) = (-1)^{\widetilde{a}_{i}\widetilde{a}_{j}}(a_{1}, a_{2}, \cdots, a_{j},a_{i}, \cdots , a_{n}).
\end{equation}
\item The generalised Jacobi identities or Jacobiators
\begin{equation}
\sum_{k+l=n-1} \sum_{(k,l)-shuffels}(-1)^{\epsilon}\left( (a_{\sigma(1)}, \cdots , a_{\sigma(k)}), a_{\sigma(k+1)}, \cdots, a_{\sigma(k+l)} \right)=0
\end{equation}
hold for all $n \geq 1$. Here $(-1)^{\epsilon}$ is a sign that arises due to the exchange of homogenous elements $a_{i} \in V$. Recall that a $(k,l)$-shuffle is a permutation of the indices $1, 2, \cdots k+l$ such that $\sigma(1) < \cdots < \sigma(k)$ and $\sigma(k+1) < \cdots < \sigma(k+l)$.
\end{enumerate}

\noindent  As the brackets are multilinear  and symmetric they are completely described by their values on even elements of $V$. Thus, considering $V$ as a manifold, the $L_{\infty}$-algebra can be written in terms of a (formal) homological vector field (elements of $V$ are identified with constant vector fields in $\Vect(V)$)

\begin{equation}
Q = Q^{i}(\xi) \frac{\partial }{\partial \xi^{i}} := \sum_{n \geq 0}\frac{1}{n!} \underbrace{(\xi, \cdots, \xi)}_{n},
\end{equation}
\noindent where $\xi$ is an even element of $V$. The Jacobiators are directly equivalent to the homological condition on $Q$. \\

\noindent It must be noted that the above definitions are shifted as compared to the original definitions of Lada \& Stasheff.  Specifically, if $V = \Pi U$ is an $L_{\infty}$-algebra (as above) then we have a series of brackets on $U$ that are antisymmetric and  even/odd for an even/odd number of arguments. Let  $x_{i} \in U$ and  we define the brackets on $U$ viz
\begin{equation}
\Pi \{x_{1}, \cdots , x_{n} \} = (-1)^{(\widetilde{x}_{1}(n-1) + \widetilde{x}_{2}(n-2)+ \cdots + \widetilde{x}_{n-1})}(\Pi x_{1}, \cdots , \Pi x_{n}).
\end{equation}
One may call $V = \Pi U$ an $L_{\infty}$-antialgebra.  However, we will refer to the bracket structures on $V$ and $U$ as  $L_{\infty}$-algebras keeping in mind the above identification.\\

\begin{definition}{Definition}
A homotopy Schouten algebra is a commutative, associate, unital algebra $\mathcal{A}$ equipped with an $L_{\infty}$-algebra structure such that the odd $n$-linear operations, known as higher Schouten brackets are multiderivations over the product:
\begin{eqnarray}
(a_{1}, a_{2}, \cdots a_{r-1}, a_{r}a_{r+1}) &=& (a_{1}, a_{2}, \cdots a_{r-1}, a_{r}) a_{r+1}\\
 \nonumber &+& (-1)^{\widetilde{a_{r}}(\widetilde{a_{1}} +\widetilde{a_{2}} + \cdots + \widetilde{a_{r-1}} +1)}a_{r}(a_{1}, a_{2}, \cdots a_{r-1},a_{r+1}),
\end{eqnarray}
with $a_{I} \in \mathcal{A}$.
\end{definition}

\noindent In order to define a homotopy Poisson algebra one needs to consider a shift in parity to keep inline with our conventions. Up to this shift, the definition carries over directly.

\begin{definition}{Definition}
A homotopy Poisson algebra is a commutative, associate, unital algebra $\mathcal{A}$ equipped with an $L_{\infty}$-algebra structure such that the  $n$-linear operations, known as higher Poisson brackets (even/odd for even/odd number of arguments) are multiderivations over the product:
\begin{eqnarray}
\{a_{1}, a_{2}, \cdots a_{r-1}, a_{r}a_{r+1}\} &=& \{a_{1}, a_{2}, \cdots a_{r-1}, a_{r}\} a_{r+1}\\
 \nonumber &+& (-1)^{\widetilde{a_{r}}(\widetilde{a_{1}} +\widetilde{a_{2}} + \cdots + \widetilde{a_{r-1}} +r)}a_{r}\{a_{1}, a_{2}, \cdots a_{r-1},a_{r+1}\},
\end{eqnarray}
with $a_{I} \in \mathcal{A}$.
\end{definition}

\noindent Following Voronov \cite{voronov-2004} it is known how to construct a series of brackets from the ``initial data"-- $\left(\mathcal{L},\pi, \Delta \right)$. Here $\mathcal{L}$ is a Lie (super)algebra equipped with a projector ($\pi^{2} = \pi$) onto an abelian subalgebra satisfying the distributivity rule $\pi[a,b] = \pi[\pi a,b] + \pi[a, \pi b]$ for all $a,b \in \mathcal{L}$.  Given an element $\Delta \in \mathcal{L}$ a series of brackets on the abelian subalgebra, $V \subset \mathcal{L}$ is defined as

\begin{equation}
(a_{1},a_{2}, \cdots,a_{n}) = \pi[\cdots[[[\Delta, a_{1} ],a_{2}],\cdots a_{n}],
\end{equation}

\noindent with $a_{i}$ in $V$.  Such brackets have the same parity as $\Delta$ and are symmetric.  The series of brackets is referred to as higher derived brackets generated by $\Delta$. A theorem due to Voronov states that  for an odd generator $\Delta \in \mathcal{L}$ the n-th Jacobiator is given by the n-th higher derived bracket generated by $\Delta^{2}$.

\begin{equation}
J^{n}(a_{1},a_{2}, \cdots,a_{n}) = \pi [\cdots[[[\Delta^{2}, a_{1} ],a_{2}],\cdots a_{n}].
\end{equation}

\noindent In particular we have that if $\Delta^{2} =0$ then the series of higher derived brackets is an $L_{\infty}$-algebra.\\

\begin{definition}{Definition}\label{Def homotopy BV algebra}
A homotopy BV-algebra is the pair $(\mathcal{A}, \Delta)$, where $\mathcal{A}$ is  a commutative, associative, unital algebra, ($C^{\infty}(M)$ for example) and $\Delta \in \End(\mathcal{A}) $ is an odd nilpotent operator. The series of odd higher brackets are given by
\begin{equation}
 [a_{1},a_{2}, \cdots,a_{n}]_{\Delta} = [\cdots[[[\Delta, a_{1} ],a_{2}],\cdots a_{n}](1),
 \end{equation}
\noindent  with $a_{i} \in V$, give an $L_{\infty}$-algebra structure on $\mathcal{A}$.
\end{definition}

\noindent One considers the Lie algebra $\mathcal{L} = \End(\mathcal{A})$ and thinks of $ V = \mathcal{A}$ as an abelian subalgebra. The projector $\pi$ is provided by the evaluation at $1$. If the generating operator $\Delta$ is a differential operator of order at most $r$ (say), then the $r+1$ th bracket is identically zero.  For further details see for example \cite{Akman:1995tm, Alfaro:1995vw,Bering:2006eb,Bering:1996kw,voronov-2004}.  \\

\noindent  The notion of homotopy ``something" algebra as used here is much more restrictive that that found elsewhere in the literature \cite{galvezcarrillo-2009,Ginzburg-1994}.   The notion used throughout this work seems very well suited to geometric considerations and suits the purposes explored here. \\

\section{Tulczyjew triples for Lie algebroids}\label{tulczyjew triples}

\noindent In this section we will construct two analogues of the classical Tulczyjew triple for Lie algebroids in the graded setting. These triples are of intrinsic interest as they appear very naturally. They also serve as the framework  for Section \ref{higher structures} where we discuss higher Poisson and Schouten structures on Lie algebroids.\\

\noindent  Given the constructions presented in this section, the ``ungraded non-super" triple
\begin{equation}
  T^{*}E \longleftarrow T^{*}E^{*} \longrightarrow TE^{*},
\end{equation}
\noindent for the Lie algebroid $E \rightarrow M$ can be recovered by careful removal of sign factors, see \cite{Mackenzie:2005,mackenzie-2002,Mackenzie:1994} for example. One can then recover the classical Tulczyjew triple for $E = TM$ directly \cite{Tulczyjew:1974}.\\

\subsection{The Tulczyjew--Schouten triple}\label{Tulczyjew--Schouten}

\noindent  Let $E \rightarrow M$ be a Lie algebroid. Consider the following commutative diagram;

\begin{diagram}[htriangleheight=30pt ]
T^{*}(\Pi E^{*})& \rTo^{\phi_{S}} & \Pi T(\Pi E^{*})[-1]\\
\dTo^{R} & \ruTo_{\psi_{S}} & \\
T^{*}(\Pi E) & &
\end{diagram}

\noindent We consider the above diagram as an analogue of the Tulczyjew triple for Lie algebroids. We refer to this triple as the \emph{Tulczyjew--Schouten triple}. The nomenclature follows form the fact that this construction uses the Schouten structure that describes the Lie algebroid explicitly. All the maps in the above diagram are double vector bundle morphisms in the category of graded manifolds.  This assertion needs explaining.\\

\noindent Let us employ natural local coordinates;

\begin{tabular}{|l ||l| }
\hline
$ T^{*}(\Pi E^{*})$ & $\{x^{A},\eta_{\alpha}, p_{A}, \pi^{\alpha} \}$ \\
$  T^{*}(\Pi E)$  & $\{x^{A},\xi^{\alpha}, p_{A}, \pi_{\alpha} \}$\\
$\Pi T(\Pi E^{*})[-1]$ &  $\{x^{A},\eta_{\alpha} , \nu^{A}, \theta_{\alpha} \}$\\
\hline
\end{tabular}\\

\noindent The parities are given by $\widetilde{x}^{A} = \widetilde{p}_{A} =\widetilde{A}$,  $\widetilde{\nu}^{A} = \widetilde{A}+1$, $\widetilde{\xi}^{\alpha} = \widetilde{\eta}_{\alpha} = \widetilde{\pi}^{\alpha} = \widetilde{\pi}_{\alpha} = \widetilde{\alpha}+1$ and $\widetilde{\theta}_{\alpha} = \widetilde{\alpha}$. The weights (relative to the natural weight on $\Pi E^{*}$) are;

\begin{tabular}{|l l  |}
\hline
 $\w(x^{A})= 0$ & $\w(\eta_{\alpha})= 1$\\
$\w(p_{A})=0$ & $\w(\pi^{\alpha})=-1$\\
$\w(\xi^{\alpha}) = -1$ & $\w(\pi_{\alpha}) = 1$\\
$\w(\nu^{A})= -1$ & $\w(\theta_{\alpha})=0$\\
\hline
\end{tabular}\\

\noindent The changes of local coordinates are given by;

\vspace{15pt}
\begin{tabular}{|l||l|}
\hline
$T^{*}(\Pi E^{*})$ &   $\overline{x}^{A}  =  \overline{x}^{A}(x)$, \hspace{5pt} $\overline{\eta}_{\alpha}  =   (T^{-1})_{\alpha}^{\:\: \beta}\eta_{\beta}$,\\
 & $\overline{p}_{A} = \left( \frac{\partial x^{B}}{\partial \overline{x}^{A}} \right)p_{B} + (-1)^{\widetilde{A}(\widetilde{\gamma}+ 1) + \widetilde{\delta}} \pi^{\delta}T_{\delta}^{\:\: \gamma} \left( \frac{\partial (T^{-1})_{\gamma}^{\:\: \alpha}}{\partial \overline{x}^{A}} \right)\eta_{\alpha}$,\\
 & $ \overline{\pi}^{\alpha} = (-1)^{\widetilde{\alpha} + \widetilde{\beta}}\pi^{\beta}T_{\beta}^{\:\: \alpha}$.\\
\hline
$T^{*}(\Pi E)$ & $ \overline{x}^{A}  =  \overline{x}^{A}(x)$, \hspace{5pt}$\overline{\xi}^{\alpha}  =   \xi^{\beta} T_{\beta}^{\:\: \alpha}$,\\
& $\overline{p}_{A} = \left( \frac{\partial x^{B}}{\partial \overline{x}^{A}} \right)p_{B} + (-1)^{\widetilde{A}(\widetilde{\gamma}+1)} \xi^{\delta}T_{\delta}^{\:\: \gamma} \left(\frac{\partial (T^{-1})_{\gamma}^{\:\: \alpha}}{\partial \overline{x}^{A}}  \right)\pi_{\alpha}$,\\
&  $\overline{\pi}_{\alpha} = (T^{-1})_{\alpha}^{\:\: \beta} \pi_{\beta}$.\\
\hline
$\Pi T(\Pi E^{*})[-1]$ & $\overline{x}^{A}  =  \overline{x}^{A}(x)$, \hspace{5pt} $\overline{\eta}_{\alpha}  =   (T^{-1})_{\alpha}^{\:\: \beta}\eta_{\beta}$, \hspace{5pt} $\overline{\nu}^{A} = \nu^{B} \left(\frac{\partial \overline{x}^{A}}{\partial x^{B}}  \right)$,\\
& $\overline{\theta}_{\alpha} = (-1)^{\widetilde{\alpha} + \widetilde{\beta}}(T^{-1})_{\alpha}^{\:\: \beta}\theta_{\beta} + \nu^{B} \left( \frac{\partial (T^{-1})_{\alpha}^{\: \: \beta}}{\partial x^{B}} \right)\eta_{\beta}$.\\
\hline
\end{tabular}\\

\noindent These transformation rules clearly show that $T^{*}(\Pi E^{*})$, $T^{*}(\Pi E)$ and $\Pi T(\Pi E^{*})[-1]$ are all double vector bundles, see Eqn. (\ref{admisiable transformations}).\\

\noindent  The algebroid structure on $E \rightarrow M$ is equivalent to a weight minus one Schouten structure on the total space of $\Pi E^{*}$. That is we have an odd function $S \in C^{\infty}( T^{*}(\Pi E^{*}))$ that satisfies the \emph{classical master equation} $\{S,S\}=0$, here the bracket is the canonical Poisson bracket. In  natural local coordinates the Schouten structure is given by

\begin{equation}
S(x,\eta,p, \pi) = \pi^{\alpha} S_{\alpha}^{A}(x) p_{A}+ \frac{1}{2!} \pi^{\alpha}\pi^{\beta}S_{\beta \alpha}^{\gamma}(x)\eta_{\gamma},
\end{equation}

\noindent with $S_{\alpha}^{A} = (-1)^{\widetilde{\alpha}} Q_{\alpha}^{A}$ and $S_{\beta \alpha}^{\gamma} = (-1)^{\widetilde{\alpha} + \widetilde{\beta}} Q_{\beta \alpha}^{\gamma}$.\\

\noindent The Schouten bracket on $C^{\infty}(\Pi E^{*})$ is given by

\begin{eqnarray}
\nonumber \SN{X, Y}_{S} &=& (-1)^{\widetilde{X}+1} \left \{ \{S,X \},Y   \right\}\\
\nonumber &=& S_{\alpha}^{A}\left( (-1)^{(\widetilde{X} +1)(\widetilde{A}+1) + \widetilde{A}\widetilde{\alpha}  } \frac{\partial X}{\partial \eta_{\alpha}} \frac{\partial Y}{\partial x^{A}}  - (-1)^{\widetilde{X}\widetilde{\alpha}} \frac{\partial X}{\partial x^{A}} \frac{\partial Y}{\partial \eta_{\alpha}}\right)\\
&-&(-1)^{\widetilde{X}\widetilde{\alpha}} S_{\alpha \beta}^{\gamma} \eta_{\gamma} \frac{\partial X}{\partial \eta_{\beta}}\frac{\partial Y}{\partial \eta_{\alpha}},
\end{eqnarray}

\noindent with $X,Y \in C^{\infty}(\Pi E^{*})$. This Schouten bracket should be thought of as the Lie algebroid generalisation of the Schouten--Nijenhuis  bracket on multivector fields over a manifold, which itself is the extension of the Lie bracket of vector fields (with shifted parity). It can be shown that the appropriate Jacobi identities are satisfied due to the condition $\{S,S\}=0$, \cite{voronov-2004,voronov-2005}. \\

\noindent Associated with the Schouten structure is the map $\phi_{S} : T^{*}(\Pi E^{*}) \rightarrow \Pi T(\Pi E^{*})[-1]$ which we will refer to as the \emph{Schouten anchor}. In natural local coordinates the Schouten anchor is given by

\begin{eqnarray} \label{schouten anchor}
\nonumber \phi_{S}^{*}(\nu^{A}) &=& \frac{\partial S}{\partial p_{A}} = \pi^{\alpha} S_{\alpha}^{A},\\
\phi_{S}^{*}(\theta_{\alpha}) &=& \frac{\partial S}{\partial \pi^{\alpha}} = S_{\alpha}^{A}p_{A} +\pi^{\beta}S_{\beta \alpha}^{\gamma}\eta_{\gamma}.
\end{eqnarray}

\noindent The Schouten anchor preserves both the parity and weight. Moreover, it is a morphism of double vector bundles. \\

\noindent It is well-known (see \cite{Mackenzie:1994,roytenberg-1999,Voronov:2001qf} for example) that there exist a canonical double vector bundle diffeomorphism  $R: T^{*}(\Pi E^{*}) \rightarrow T^{*}(\Pi E )$. This morphism is independent of any Lie algebroid structure on $E$ and is a symplectomorphism between the canonical even symplectic structures. This is easily verified in natural local coordinates. Comparing the above with the transformation rules for the coordinates on $T^{*}(\Pi E^{*})$ and $T^{*}(\Pi E)$  we see that the required diffeomorphism is given by

\begin{equation}
R^{*}(\pi_{\alpha}) = \eta_{\alpha},  \hspace{35pt} R^{*}(\xi^{\alpha}) = (-1)^{\widetilde{\alpha}}\pi^{\alpha}.
\end{equation}

\noindent  To see that $R$ is indeed a symplectomorphism,  note that the canonical even  symplectic structure on $T^{*}(\Pi E^{*})$ is given by $\omega_{T^{*}(\Pi E^{*})} =  dp_{A}dx^{A} + d\pi^{\alpha}d \eta_{\alpha}$ and on $T^{*}(\Pi E)$ it is given by $\omega_{T^{*}(\Pi E)} = dp_{A}dx^{A} + d\pi_{\alpha}d\xi^{\alpha}$. Thus, $R^{*}\omega_{T^{*}(\Pi E )} = \omega_{T^{*}(\Pi E^{*})}$. \\

\noindent We then define the \emph{Tulczyjew--Schouten morphism} as the composition of the inverse of the canonical double vector bundle morphism and the Schouten anchor:
 \begin{equation}
 \psi_{S} = \phi_{S} \circ R^{-1} : T^{*}(\Pi E) \rightarrow \Pi T(\Pi E^{*})[-1].
 \end{equation}

\noindent  Then in natural coordinates we have

\begin{eqnarray}
\nonumber \psi_{S}^{*}(\eta_{\alpha}) &=& \pi_{\alpha}, \\
\nonumber  \psi_{S}^{*}(\nu^{A}) &=& (-1)^{\widetilde{A}} \xi^{\alpha} S_{\alpha}^{A},\\
\psi_{S}^{*}(\theta_{\alpha}) &=& S_{\alpha}^{A}p_{A} +  (-1)^{\widetilde{\beta}}\xi^{\beta}S_{\beta \alpha}^{\gamma} \pi_{\gamma}.
\end{eqnarray}

\noindent One can also discuss Lie algebroids in the language of gauge systems. Consider \newline
$\{T^{*}(\Pi E^{*}), \{\bullet, \bullet \}_{T^{*}(\Pi E^{*})}, S  \}$ as  a gauge system. Equivalently, the information in the definition of a Lie algebroid can be encoded in the gauge system $\{T^{*}(\Pi E), \{ \bullet, \bullet \}_{T^{*}(\Pi E)} , H_{Q} \}$. Here $H_{Q} \in C^{\infty}(T^{*}(\Pi E))$ is the \emph{linear Hamiltonian}  associated with the homological vector field $Q \in \Vect(\Pi E)$. Note that the homological condition on $Q$ becomes the \emph{master equation} $\{H_{Q}, H_{Q} \}_{T^{*}(\Pi E)}=0$. In natural local coordinates we have;

\begin{equation}
H_{Q} = \xi^{\alpha}Q_{\alpha}^{A}p_{A} + \frac{1}{2} \xi^{\alpha} \xi^{\beta}Q_{\beta \alpha}^{\gamma}\pi_{\gamma}.
\end{equation}

\begin{theorem}{Theorem}
The diffeomorphism $R: T^{*}(\Pi E^{*}) \rightarrow T^{*}(\Pi E)$ is a morphism in the category of gauge systems.
\end{theorem}

\begin{proof}
The proof comes in two stages;
\begin{enumerate}
\item As $R$ is symplectomorphism we have $R^{*}\{ f,g \}_{T^{*}(\Pi E)} =  \{ R^{*}f, R^{*}g \}_{T^{*}(\Pi E^{*})}$ for all \newline $f,g \in C^{\infty}(T^{*}(\Pi E))$.
\item The fact that $R^{*}H_{Q} = S$ can be easily verified via local coordinates.
\end{enumerate}
Thus, $R$ is a morphism between gauge systems.
\end{proof}

\begin{example}{Example}
\noindent Consider the vector bundle $E = TN$.  The corresponding diagram is given by:

\begin{diagram}[htriangleheight=30pt ]
T^{*}(\Pi T^{*}N)& \rTo^{\phi_{S}} & \Pi T(\Pi T^{*}N)[-1]\\
\dTo^{R} & \ruTo_{\psi_{S}} & \\
T^{*}(\Pi TN) & &
\end{diagram}

\noindent Let us work in the following natural local coordinates;

\begin{tabular}{|l ||l| }
\hline
$T^{*}(\Pi T^{*}N)$ & $\{x^{A},x^{*}_{A}, p_{A}, \pi^{A} \}$\\
$ T^{*}(\Pi TN)$  & $\{x^{A},\xi^{A}, p_{A}, \pi_{A} \}$\\
$\Pi T(\Pi T^{*}N)[-1]$ &  $\{x^{A},x^{*}_{A} , \nu^{A}, \theta_{A} \}$\\
\hline
\end{tabular}\\

\noindent For this example we have $S_{B}^{A} = (-1)^{\widetilde{A}} \delta_{B}^{A}$ and $S_{AB}^{C}=0$ for the canonical Schouten structure. Note that $C^{\infty}(\Pi T^{*}M)$  is defined to be the space of multivector fields over $M$ and the  Schouten bracket  is  the Schouten--Nijenhuis bracket. Then we see that

\begin{equation}
\phi_{S}^{*}(\nu^{A}) = (-1)^{\widetilde{A}} \pi^{A}, \hspace{30pt}  \phi^{*}_{S}(\theta_{A}) = (-1)^{\widetilde{A}}p_{A},
\end{equation}

\begin{equation}
R^{*}(\pi_{A}) = x^{*}_{A}, \hspace{30pt} R^{*}(\xi^{A}) = (-1)^{\widetilde{A}} \pi^{A},
\end{equation}

\begin{equation}
\psi_{S}^{*}(x^{*}_{A}) = \pi_{A}, \hspace{10pt} \psi_{S}^{*}(\nu^{A}) = \xi^{A}, \hspace{10pt} \psi_{S}^{*}(\theta_{A}) = (-1)^{\widetilde{A}}p_{A}.
\end{equation}

\noindent The above identifications can easily be verified  independently of the discussion about Lie algebroids  by examining the local coordinates and there transformation rules.\\

\noindent It must be noted that for this case  all the identifications are diffeomorphisms. In particular the diffeomorphism $\psi_{S}$ is regarded as  the inverse of the appropriately graded Tulczyjew diffeomorphism

\begin{equation}
\alpha = \psi_{S}^{-1} : \Pi T(\Pi T^{*}N)[-1] \longrightarrow T^{*}(\Pi TN).
\end{equation}

\noindent In this example we have $S = (-1)^{\widetilde{A}} \pi^{A}p_{A}$ and $H_{Q} =  \xi^{A}p_{A}$, the Hamiltonian associated with  the de Rham differential on $N$.  It is a straight forward to verify that $R^{*}(H_{Q}) = S$. \\

\noindent More importantly, as  both $\phi_{S}$ and $\psi_{S}$ are invertible one can pullback the (equivalent) canonical symplectic structures  to $\Pi T(\Pi T^{*}N)[-1]$. This even symplectic structure we refer to as the \emph{canonical Koszul--Poisson structure}.  It provides an even Poisson bracket on the space of differential forms over the Schouten (odd symplectic) manifold $\Pi T^{*}N$.

\begin{eqnarray}
\nonumber \omega_{\Pi T(\Pi T^{*}N)} &:=& (\phi^{-1}_{S})^{*}\omega_{T^{*}(\Pi T^{*}N)} =  (\psi^{-1}_{S})^{*}\omega_{T^{*}(\Pi TN)}\\
&=& (-1)^{\widetilde{A}} d\theta_{A}dx^{A} + (-1)^{\widetilde{A}} d\nu^{A}dx^{*}_{A}.
\end{eqnarray}

\end{example}

\begin{example}{Example}
\noindent In the other extreme we have Lie algebras. Consider a vector space $\mathfrak{g}$ equipped with a Lie bracket $[,]: \mathfrak{g} \otimes \mathfrak{g} \rightarrow \mathfrak{g}$. One thinks of the Lie algebra as a Lie algebroid over a point. If we pick a basis $\{x_{\alpha} \}$ the bracket can be written as $[x_{\alpha}, x_{\beta}] = (-1)^{\widetilde{\beta}} Q_{\alpha \beta}^{\gamma}x_{\gamma}$.\\

\noindent For this example we need to consider the manifolds equipped with the local coordinates: \\
\begin{tabular}{|l ||l| }
\hline
$T^{*}(\Pi \mathfrak{g}^{*})$ & $\{\eta_{\alpha}, \pi^{\alpha} \}$\\
$ T^{*}(\Pi \mathfrak{g})$  & $\{\xi^{\alpha},  \pi_{\alpha} \}$\\
$\Pi T(\Pi \mathfrak{g}^{*})[-1]$ &  $\{\eta_{\alpha} ,  \theta_{\alpha} \}$\\
\hline
\end{tabular}\\

\noindent The Lie algebra structure on $\mathfrak{g}$ is equivalent to a weight minus one Schouten structure on $\Pi \mathfrak{g}^{*}$. In local coordinates   the structure is given by $S = (-1)^{\widetilde{\alpha} + \widetilde{\beta}}\frac{1}{2!} \pi^{\alpha} \pi^{\beta} Q_{\beta \alpha}^{\gamma}\eta_{\gamma}$. The associated odd brackets are known in the literature as the Lie--Schouten brackets.\\

\noindent We then have the following diagram

\begin{diagram}[htriangleheight=30pt ]
T^{*}(\Pi \mathfrak{g}^{*})& \rTo^{\phi_{S}} & \Pi T(\Pi \mathfrak{g}^{*})[-1]\\
\dTo^{R} & \ruTo_{\psi_{S}} & \\
T^{*}(\Pi \mathfrak{g}) & &
\end{diagram}

\noindent Employing local coordinates we have

\begin{equation}
\phi^{*}_{S}(\theta_{\alpha}) = (-1)^{\widetilde{\alpha} + \widetilde{\beta}}\pi^{\beta}Q_{\beta \alpha}^{\gamma} \eta_{\gamma},
\end{equation}
\begin{equation}
\psi^{*}_{S}(\eta_{\alpha}) = \pi_{\alpha} , \hspace{25pt} \psi^{*}_{S}(\theta_{\alpha}) = (-1)^{\widetilde{\alpha}}\xi^{\beta} Q_{\beta \alpha}^{\gamma}\pi_{\gamma}.
\end{equation}
\end{example}

\newpage

 \subsection{The Tulczyjew--Poisson triple}\label{Tulczyjew--Poisson}

\noindent In the previous subsection Lie algebroids were discussed in terms of double vector bundles and Schouten structures. It is clear that a parallel ``even" description in terms of  Poisson structures exists. We proceed to describe this in less detail as it parallels the previous section closely.\\

\noindent The corresponding commutative diagram is given by;

\begin{diagram}[htriangleheight=30pt ]
\Pi T^{*}( E^{*})& \rTo^{\phi_{P}} & \Pi T( E^{*})[-1]\\
\dTo^{R} & \ruTo_{\psi_{P}} & \\
\Pi T^{*}(\Pi E) & &
\end{diagram}

\noindent We refer to this triple as the \emph{Tulczyjew-Poisson triple}. The morphisms in the above diagram are double vector bundle morphisms in the category of graded manifolds. Let us employ natural local coordinates;

\begin{tabular}{|l ||l| }
\hline
$\Pi T^{*}(E^{*})$ & $\{x^{A},e_{\alpha}, x^{*}_{A}, e_{*}^{\alpha} \}$\\
$ \Pi T^{*}(\Pi E)$  & $\{x^{A},\eta^{\alpha}, x^{*}_{A}, \eta_{\alpha}^{*} \}$\\
$\Pi T(E^{*})[-1]$ &  $\{x^{A},e_{\alpha} , \xi^{A}, \theta_{\alpha} \}$\\
\hline
\end{tabular}\\

\noindent The  respective parities are given by $\widetilde{x}^{A} = \widetilde{A}$, $\widetilde{e}^{\alpha} = \widetilde{\eta}^{*}_{\alpha}= \widetilde{\theta}_{\alpha} = \widetilde{\alpha}$, $ \widetilde{x}^{*}_{A} = \widetilde{\xi}^{A}=\widetilde{A}+1$ and $\widetilde{\eta}^{\alpha}= \widetilde{e}_{*}^{\alpha} = \widetilde{\alpha}+1$. The weights are;

\begin{tabular}{|l l|}
\hline
$\w(x^{A})= 0$ & $\w(e_{\alpha})= 1$\\
$\w(x^{*}_{A}) = 0 $ & $\w(e_{*}^{\alpha}) = -1$\\
$\w(\nu^{A})= -1$ & $\w(\nu^{*}_{\alpha})=1$\\
$\w(\xi^{A})= -1$ & $\w(\theta_{\alpha})=0$\\
\hline
\end{tabular}\\

\noindent The changes of coordinates are almost identical to the case discussed in the previous section with subtle sign changes.

\begin{tabular}{|l||l|}
\hline
$\Pi T^{*}( E^{*})$ &   $\overline{x}^{A}  =  \overline{x}^{A}(x)$, \hspace{5pt} $\overline{e}_{\alpha}  =   (T^{-1})_{\alpha}^{\:\: \beta}e_{\beta}$,\\
 & $\overline{x}^{*}_{A} = \left( \frac{\partial x^{B}}{\partial \overline{x}^{A}} \right)x^{*}_{B} - (-1)^{\widetilde{A}(\widetilde{\gamma}+ 1) + \widetilde{\delta}} e_{*}^{\delta}T_{\delta}^{\:\: \gamma} \left( \frac{\partial (T^{-1})_{\gamma}^{\:\: \alpha}}{\partial \overline{x}^{A}} \right)e_{\alpha}$,\\
 & $ \overline{e}_{*}^{\alpha} = e_{*}^{\beta}T_{\beta}^{\:\: \alpha}$.\\
\hline
$\Pi T^{*}(\Pi E)$ & $ \overline{x}^{A}  =  \overline{x}^{A}(x)$, \hspace{5pt}$\overline{\eta}^{\alpha}  =   \eta^{\beta} T_{\beta}^{\:\: \alpha}$,\\
& $\overline{x}^{*}_{A} = \left( \frac{\partial x^{B}}{\partial \overline{x}^{A}} \right)x^{*}_{B} + (-1)^{\widetilde{A}(\widetilde{\gamma}+1)} \eta^{\delta}T_{\delta}^{\:\: \gamma} \left(\frac{\partial (T^{-1})_{\gamma}^{\:\: \alpha}}{\partial \overline{x}^{A}}  \right)\eta^{*}_{\alpha}$,\\
&  $\overline{\eta}^{*}_{\alpha} = (T^{-1})_{\alpha}^{\:\: \beta} \eta^{*}_{\beta}$.\\
\hline
$\Pi T(E^{*})[-1]$ & $\overline{x}^{A}  =  \overline{x}^{A}(x)$, \hspace{5pt} $\overline{e}_{\alpha}  =   (T^{-1})_{\alpha}^{\:\: \beta}e_{\beta}$, \hspace{5pt} $\overline{\xi}^{A} = \xi^{B} \left(\frac{\partial \overline{x}^{A}}{\partial x^{B}}  \right)$,\\
& $\overline{\theta}_{\alpha} = (-1)^{\widetilde{\alpha} + \widetilde{\beta}}(T^{-1})_{\alpha}^{\:\: \beta}\theta_{\beta} + \xi^{B} \left( \frac{\partial (T^{-1})_{\alpha}^{\: \: \beta}}{\partial x^{B}} \right)e_{\beta}$.\\
\hline
\end{tabular}\\

\noindent The canonical double vector bundle diffeomorphism $R:\Pi T^{*}(E^{*}) \rightarrow \Pi T^{*}(\Pi E)$ is given by

\begin{equation}
R^{*}(\eta^{\alpha}) = e_{*}^{\alpha}, \hspace{30pt} R^{*}(\eta_{\alpha}^{*}) = -e_{\alpha}.
\end{equation}

\noindent The canonical double vector bundle diffeomorphism is a symplectomorphism between the canonical odd symplectic structures. The canonical odd symplectic structure on $\Pi T^{*}(E^{*})$ is  given by $\omega_{\Pi T^{*}(E^{*})} = (-1)^{\widetilde{A}+1} dx^{*}_{A}dx^{A} + (-1)^{\widetilde{\alpha}+1} de_{*}^{\alpha}de_{\alpha}$ and on $\Pi T^{*}(\Pi E)$ it is given by  $\omega_{\Pi T^{*}(\Pi E)} = (-1)^{\widetilde{A}+1} dx^{*}_{A}dx^{A} + (-1)^{\alpha} d \eta_{\alpha}^{*}d \eta^{\alpha}$.  It is then a simple exercise to see that  $R^{*}(\omega_{\Pi T^{*}(\Pi E)}) = \omega_{\Pi T^{*}(E^{*})}$.\\

\noindent In these natural coordinates the weight minus one Poisson structure describing the Lie algebroid $E \rightarrow M$ is given by

\begin{equation}
P(x,e,x^{*}, e_{*}) = e_{*}^{\alpha}P_{\alpha}^{A}(x)x^{*}_{A} + \frac{1}{2} e_{*}^{\alpha}e_{*}^{\beta}P_{\beta \alpha}^{\gamma}(x) e_{\gamma}.
\end{equation}

\noindent Here $P_{\alpha}^{A} =  Q_{\alpha}^{A}$ and $P_{\beta \alpha}^{\gamma} = - Q_{\beta \alpha}^{\gamma}$.\\

\noindent The Poisson bracket on $C^{\infty}(E^{*})$ is given by

\begin{eqnarray}
\nonumber \{F,G \}_{P} &=& (-1)^{\widetilde{F}+1} \SN{ \SN{P,F},G}\\
\nonumber &=&  P_{\alpha}^{A}\left((-1)^{\widetilde{F}\widetilde{A} + \widetilde{A}\widetilde{\alpha}} \frac{\partial F}{\partial e_{\alpha}}\frac{\partial G}{\partial x^{A}} - (-1)^{\widetilde{F}\widetilde{\alpha}}\frac{\partial F}{\partial x^{A}}\frac{\partial G}{\partial e_{\alpha}}   \right)\\
&+& (-1)^{\widetilde{F}\widetilde{\alpha} + \widetilde{\alpha}} P_{\beta \alpha}^{\gamma}e_{\gamma}\frac{\partial F}{\partial e_{\alpha}}\frac{\partial G}{\partial e_{\beta}},
\end{eqnarray}

\noindent with $F,G \in C^{\infty}(E^{*})$. This Poisson bracket should be thought of as the Lie algebroid generalisation of the Poisson bracket on contravariant symmetric tensors over a manifold (i.e. functions on the cotangent bundle).  The appropriate Jacobi identities are satisfied due to the condition $\SN{P,P}=0$.\\

\noindent Associated with the Poisson structure is the map $\phi_{P}: \Pi T^{*}(E^{*}) \rightarrow \Pi T(E^{*})[-1]$ which we will refer to as the \emph{Poisson anchor}. In natural local coordinates the Poisson anchor is given by

\begin{eqnarray}
\nonumber \phi_{P}^{*}(\xi^{A}) &=& (-1)^{\widetilde{A}+1}\frac{\partial P}{\partial x^{*}_{A}} = e^{\alpha}_{*} P_{\alpha}^{A},\\
\phi_{P}^{*}(\theta_{\alpha}) &=& (-1)^{\widetilde{\alpha}+1} \frac{\partial P}{\partial e^{\alpha}_{*}} = (-1)^{\widetilde{\alpha}+1} P_{\alpha}^{A}x^{*}_{A} + (-1)^{\widetilde{\alpha} + 1}e^{\beta}_{*}P_{\beta \alpha}^{\gamma} e_{\gamma}.
\end{eqnarray}

\noindent Then we define the \emph{Tulczyjew--Poisson morphism} as the composition of the Poisson anchor and the inverse of the canonical double vector bundle morphism:
\begin{equation}
\psi_{P} = \phi_{P} \circ R^{-1} : \Pi T^{*}(\Pi E)  \rightarrow \Pi T(E^{*})[-1].
\end{equation}

\noindent In natural local coordinates we have;

\begin{eqnarray}
\nonumber \psi_{P}^{*}(e_{\alpha}) &=& - \eta^{*}_{\alpha},\\
\nonumber \psi_{P}^{*}(\xi^{A}) &=&\eta^{\alpha}P_{\alpha}^{A},\\
\psi_{P}^{*}(\theta_{\alpha}) &=& (-1)^{\widetilde{\alpha} +1} P_{\alpha}^{A}x^{*}_{A} + (-1)^{\widetilde{\alpha}} \eta^{\beta}P_{\beta \alpha}^{\gamma} \eta^{*}_{\gamma}.
\end{eqnarray}

\noindent Again, one can think in terms of gauge systems. The Lie algebroid structure on $E \rightarrow M$ is equivalent to the gauge system $\{\Pi T^{*}(E^{*}), \SN{\bullet, \bullet}_{\Pi T^{*}(E^{*})}, P \}$. Equivalently, one has the gauge system $\{ \Pi T^{*}(\Pi E), \SN{\bullet, \bullet}_{\Pi T^{*}(\Pi E)}, X_{Q} \}$. Where the homological potential\newline $X_{Q} = e_{*}^{\alpha}Q_{\alpha}^{A}x^{*}_{A} + \frac{1}{2}\eta^{\alpha}\eta^{\beta}Q_{\beta \alpha}^{\gamma} \eta^{*}_{\gamma} \in C^{\infty}(\Pi T^{*}(\Pi E))$ is the one-vector associated with the homological vector field $Q$. Note this one-vector is even. The homological condition on $Q$ is then equivalent to the \emph{master equation}
$\SN{X_{Q}, X_{Q}}_{\Pi T^{*}(\Pi E)}=0$.\\

\begin{theorem}{Theorem}
The canonical double vector bundle morphism $R :\Pi T^{*}(E^{*}) \rightarrow \Pi T^{*}(\Pi E) $ is a morphism in the category of gauge systems.
\end{theorem}

\begin{proof}
The proof follows almost identically to the Schouten case.
\end{proof}

\begin{example}{Example}Consider the vector bundle $E = TN$.  It is clear that the relevant diagram is given by

\begin{diagram}[htriangleheight=30pt ]
\Pi T^{*}( T^{*}N)& \rTo^{\phi_{P}} & \Pi T( T^{*}N)[-1]\\
\dTo^{R} & \ruTo_{\psi_{P}} & \\
\Pi T^{*}(\Pi TN) & &
\end{diagram}

\noindent Let us work in the following natural local coordinates;

\begin{tabular}{|l ||l| }
\hline
$\Pi T^{*}( T^{*}N)$ & $\{x^{A},p_{A}, x^{*}_{A}, p_{*}^{A} \}$\\
$ \Pi T^{*}(\Pi TN)$  & $\{x^{A},\eta^{A}, x^{*}_{A}, \eta^{*}_{A} \}$\\
$\Pi T( T^{*}N)[-1]$ &  $\{x^{A},p_{A} , \xi^{A}, \theta_{A} \}$\\
\hline
\end{tabular}\\

\noindent For this example we have $P_{A}^{B}= \delta_{A}^{B}$. We see that

\begin{equation}
\phi_{P}^{*}(\xi^{A}) =  p_{*}^{A}, \hspace{30pt}  \phi^{*}_{P}(\theta_{A}) = (-1)^{\widetilde{A}+1}x^{*}_{A},
\end{equation}

\begin{equation}
R^{*}(\eta^{A}) = p^{*}_{A}, \hspace{30pt} R^{*}(\eta^{*}_{A}) = - p_{A},
\end{equation}

\begin{equation}
\psi_{P}^{*}(p_{A}) = -\eta^{*}_{A}, \hspace{10pt} \psi_{P}^{*}(\xi^{A}) = \eta^{A}, \hspace{10pt} \psi_{P}^{*}(\theta_{A}) = (-1)^{\widetilde{A}+1}x^{*}_{A}.
\end{equation}

\noindent Similarly to the case examined in the previous section, the  diffeomorphism $\psi_{P}$ is regarded as  the inverse of the appropriately graded Tulczyjew diffeomorphism

\begin{equation}
\alpha = \psi_{P}^{-1} : \Pi T( T^{*}N)[-1] \longrightarrow \Pi T^{*}(\Pi TN).
\end{equation}

\noindent In this example we have $P= p_{*}^{A}x^{*}_{A}$ and $X_{Q} = \eta^{A}x^{*}_{A}$ (the one-vector associated with the de Rham differential). It is straight forward to verify that $ R^{*}(X_{Q}) =P$.\\

\noindent Furthermore,  as  both $\phi_{P}$ and $\psi_{P}$ are invertible one can pullback the (equivalent) canonical symplectic structures  to $\Pi T( T^{*}N)[-1]$. This odd symplectic structure we refer to as the \emph{canonical Koszul--Schouten structure}.  It provides a Schouten  bracket on the space of differential forms over the Poisson (even symplectic) manifold $T^{*}N$;

\begin{eqnarray}
\nonumber \omega_{\Pi T(T^{*}N)[-1]} &=& (\phi_{P}^{-1})^{*} \omega_{\Pi T^{*}(T^{*}N)} = (\psi_{P}^{-1})^{*}\omega_{\Pi T^{*}(\Pi TN)}\\
&=& d\theta_{A} dx^{A} + (-1)^{\widetilde{A}+1} dp_{A}d \xi^{A}.
\end{eqnarray}
\end{example}

\begin{example}{Example}
\noindent Consider a vector space $\mathfrak{g}$ equipped with a Lie bracket $[,]: \mathfrak{g} \otimes \mathfrak{g} \rightarrow \mathfrak{g}$. One thinks of the Lie algebra as a Lie algebroid over a point. If we pick a basis $\{x_{\alpha} \}$ the bracket can be written as $[x_{\alpha}, x_{\beta}] = (-1)^{\widetilde{\beta}} Q_{\alpha \beta}^{\gamma}x_{\gamma}$.\\

\noindent Let us work in natural local coordinates:

\begin{tabular}{|l ||l| }
\hline
$\Pi T^{*}( \mathfrak{g}^{*})$ & $\{e_{\alpha}, e_{*}^{\alpha} \}$\\
$ \Pi T^{*}(\Pi \mathfrak{g})$  & $\{\eta^{\alpha},  \eta^{*}_{\alpha} \}$\\
$\Pi T( \mathfrak{g}^{*})[-1]$ &  $\{e_{\alpha} ,  \theta_{\alpha} \}$\\
\hline
\end{tabular}\\

\noindent A Lie algebra structure on  $\mathfrak{g}$ is equivalent to a weight minus one Poisson structure on $\mathfrak{g}^{*}$. In these coordinates this structure is given by $P = - \frac{1}{2}e_{*}^{\alpha} e_{*}^{\beta} Q_{\beta \alpha}^{\gamma}e_{\gamma}$. The corresponding Poisson brackets are known in the literature as Poisson--Lie brackets.   The relevant triple is given by the diagram:

\begin{diagram}[htriangleheight=30pt ]
\Pi T^{*}( \mathfrak{g}^{*})& \rTo^{\phi_{P}} & \Pi T( \mathfrak{g}^{*})[-1]\\
\dTo^{R} & \ruTo_{\psi_{P}} & \\
\Pi T^{*}(\Pi \mathfrak{g}) & &
\end{diagram}

 \noindent Employing local coordinates we have

\begin{equation}
\phi^{*}_{P}(\theta_{\alpha}) = (-1)^{\widetilde{\alpha}} e^{\beta}_{*} Q_{\beta \alpha}^{\gamma} e_{\gamma},
\end{equation}
\begin{equation}
\psi^{*}_{P}(e_{\alpha}) = - \eta_{\alpha}^{*} , \hspace{25pt} \psi^{*}_{P}(\theta_{\alpha}) = (-1)^{\widetilde{\alpha}+1}\eta^{\beta} Q_{\beta \alpha}^{\gamma}\eta_{\gamma}^{*}.
\end{equation}
\end{example}

\section{Higher Poisson and higher Schouten structures}\label{higher structures}
\noindent Higher Poisson and higher Schouten structures on (super)manifolds were first described by Voronov \cite{voronov-2004,voronov-2005} in the context of his higher derived bracket formalism. Recall than a higher Poisson structure on a manifold $M$ is understood as an even function $P$ on the total space of $\Pi T^{*}M$  such that it self-commutes with respect to the canonical Schouten structure: $\SN{P,P}=0$. To such a structure one associates a series of higher Poisson brackets between functions on $M$ such that $C^{\infty}(M)$ becomes a homotopy Poisson algebra. That is $C^{\infty}(M)$ becomes an $L_{\infty}$-algebra (suitably ``superised" so brackets with an even/odd number of arguments are even/odd ) such that the brackets act as multiderivations over the supercommutative product of functions.\\

\noindent  Similarly, a higher Schouten structure on a manifold $M$ is understood as an odd function $S$ on the total space of $T^{*}M$ such that it self-commutes with respect to the canonical poisson structure: $\{S,S\}=0 $. To such a structure one associates a series  of higher Schouten brackets between functions on $M$ such that $C^{\infty}(M)$ becomes a homotopy Schouten algebra.

\subsection{Higher Poisson and higher Schouten structures on Lie algebroids}\label{higher Poisson and Schouten}

\noindent In this subsection we define and make initial study of higher Poisson and higher Schouten structures on Lie algebroids. We do this in analogy with Voronov's higher Poisson and Schouten structures on (super)manifolds \cite{voronov-2004,voronov-2005}. (Also see \cite{khudaverdian-2008}).

\begin{definition}{Definition}
 Let  $E \rightarrow M$ be a Lie algebroid. A higher Poisson structure on a Lie algebroid is defined to be an even function $\mathcal{P} \in C^{\infty}(\Pi E^{*})$ such that $\SN{\mathcal{P}, \mathcal{P}}_{S}=0$. Similarly, a higher Schouten structure on a Lie algebroid is defined to be an odd function $\mathcal{S} \in C^{\infty}( E^{*})$ such that $\{\mathcal{S}, \mathcal{S}\}_{P} =0$.
\end{definition}

\begin{remark}
Note we have no condition on the weight of the higher Poisson or Schouten structure. For example, the (classical) Poisson structures on Lie algebroids are recognised as  weight two higher Poisson structures.
\end{remark}

\noindent Associated with the higher Poisson structure is a homotopy Poisson algebra on $C^{\infty}(M)$. The series of higher Poisson  brackets is provided by Voronov's higher derived bracket formulism as

\begin{equation}
\{ f_{1}, f_{2}, \cdots, f_{r} \}_{\mathcal{P}} =\left. \SN{\cdots \SN{\SN{\mathcal{P},f_{1}}_{S},f_{2} }_{S}, \cdots , f_{r}   }_{S}\right|_{M},
\end{equation}

\noindent with $f_{I} \in C^{\infty}(M)$.\\

\noindent Similarly, associated with the higher Schouten structure is a homotopy Schouten algebra on $C^{\infty}(M)$. The series of higher Schouten brackets are given by

\begin{equation}
(f_{1},f_{2}, \cdots, f_{r})_{\mathcal{S}} = \left.  \{\cdots \{\{ \mathcal{S},f_{1} \}_{P}, f_{2}\}_{P},\cdots, f_{r}  \}_{P}\right|_{M}.
\end{equation}

\noindent with $f_{I} \in C^{\infty}(M)$.\\

\noindent It is well-known that for a Poisson manifold the cotangent bundle comes equipped with the structure of a Lie algebroid.  As we shall see, this  also extends to the case of higher Poisson and higher Schouten structures on Lie algebroids with the proviso that we consider $L_{\infty}$-algebroids.  Recall that an $L_{\infty}$-algebroid structure on a vector bundle $F \rightarrow M$ is defined to be a homological vector field (inhomogeneous in weight) on the total space of $\Pi F$.

\begin{theorem}{Theorem} Let $E \rightarrow M$ be a Lie algebroid.
 \begin{enumerate}
 \item If $E$ comes equipped with a higher Poisson structure $\mathcal{P}$, then $E^{*} \rightarrow M$ is  an $L_{\infty}$-algebroid.
\item If $E$ comes equipped with a higher Schouten structure $\mathcal{S}$, then $\Pi E^{*} \rightarrow M$ is  an $L_{\infty}$-algebroid.
\end{enumerate}
\end{theorem}

\begin{proof}
 We provide a proof  by explicitly finding the homological vector fields (inhomogeneous in weight) on $\Pi E^{*}$ and $E^{*}$ corresponding to the $L_{\infty}$-algebroid structures.\\

\begin{enumerate}
\item Let $S \in C^{\infty}(T^{*}(\Pi E^{*}))$ be the weight minus one Schouten structure describing the Lie algebroid $E \rightarrow M$. Then consider
\begin{eqnarray}
 \mathcal{H}_{\mathcal{P}} &=& \{ S, \mathcal{P} \}_{T^{*}(\Pi E^{*})}\\
\nonumber &=& \left( \frac{\partial \mathcal{P}}{\partial \eta_{\alpha}}S^{A}_{\alpha} \right)p_{A} + \left( (-1)^{\widetilde{A}(\widetilde{\alpha}+1)}\frac{\partial \mathcal{P}}{\partial x^{A}}S^{A}_{\alpha} + \frac{\partial \mathcal{P}}{\partial \eta_{\beta}} S^{\gamma}_{\beta\alpha}\eta_{\gamma}  \right)\pi^{\alpha} \in C^{\infty}(T^{*}(\Pi E^{*})).
\end{eqnarray}
Note that direct application of the Jacobi identity produces $\{S, \mathcal{H}_{\mathcal{P}}  \}_{T^{*}(\Pi E^{*})}=0.$ Furthermore, from the definition of the Schouten bracket we see that \newline $\SN{\mathcal{P}, \mathcal{P}}_{S}= \{\mathcal{P}, \mathcal{H}_{\mathcal{P}}  \}_{T^{*}(\Pi E^{*})}$. The direct application of the Jacobi identities produces
\begin{equation}
\{\mathcal{H}_{\mathcal{P}}, \mathcal{H}_{\mathcal{P}}  \}_{T^{*}(\Pi E^{*})} =0.
\end{equation}
Thus, $\mathcal{H}_{\mathcal{P}}$ is interpreted as a \emph{linear Hamiltonian} associated with a homological vector field on $\Pi E^{*}$:
\begin{eqnarray}
Q_{\mathcal{P}} &=& \left( \frac{\partial \mathcal{P}}{\partial \eta_{\alpha}}S^{A}_{\alpha} \right)\frac{\partial}{\partial x^{A}}\\
\nonumber &+& \left( (-1)^{\widetilde{A}(\widetilde{\alpha}+1)}\frac{\partial \mathcal{P}}{\partial x^{A}}S^{A}_{\alpha} + \frac{\partial \mathcal{P}}{\partial \eta_{\beta}} S^{\gamma}_{\beta\alpha}\eta_{\gamma}  \right) \frac{\partial}{\partial \eta_{\alpha}} \in \Vect(\Pi E^{*}).
\end{eqnarray}
Thus, the vector bundle $E^{*}\rightarrow M$ is an $L_{\infty}$-algebroid. Furthermore, we see that the homological vector field is  the Hamiltonian vector field associated with the higher Poisson structure; $Q_{\mathcal{P}}= - \SN{\mathcal{P}, \bullet}_{S}$.
\item Let $P\in C^{\infty}(\Pi E^{*})$ be the wight minus one Poisson structure describing the Lie algebroid $E \rightarrow M$. Then consider
\begin{eqnarray}
 \mathcal{H}_{\mathcal{S}} &=& \SN{P, \mathcal{S}}_{\Pi T^{*}(E^{*})} =  \left( \frac{\partial \mathcal{S}}{\partial e_{\alpha}}P_{\alpha}^{A} \right)x^{*}_{A} \\
\nonumber &+& (-1)^{\widetilde{\alpha}+1} \left(\frac{\partial \mathcal{S}}{\partial e_{\beta}}P_{\beta \alpha}^{\gamma}e_{\gamma} + (-1)^{\widetilde{\alpha} (\widetilde{A}+1)} \frac{\partial \mathcal{S}}{\partial x^{A}} P_{\alpha}^{A}  \right)e_{*}^{\alpha} \in C^{\infty}(\Pi T^{*}(E^{*})).
\end{eqnarray}
Note that direct application of the Jacobi identity produces $\SN{P, \mathcal{H}_{\mathcal{S}}}_{\Pi T^{*}(E^{*})} =0$. Furthermore, from the definition of the Poisson bracket we see that \newline $\{\mathcal{S} , \mathcal{S}\}_{P} = - \SN{\mathcal{S}, \mathcal{H}_{\mathcal{S}}}_{\Pi T^{*}(E^{*})}$. Then direct application of the Jacobi identities produces
\begin{equation}
\SN{\mathcal{H}_{\mathcal{S}} , \mathcal{H}_{\mathcal{S}}}_{\Pi T^{*}(E^{*})} = 0.
\end{equation}
Thus,  via the \emph{odd isomorphism} one associates a homological vector field with $\mathcal{H}_{\mathcal{S}}$:
\begin{eqnarray}
Q_{\mathcal{S}} &=& \left( \frac{\partial \mathcal{S}}{\partial e_{\alpha}}P_{\alpha}^{A} \right)\frac{\partial}{\partial x^{A}}\\
 \nonumber &+& (-1)^{\widetilde{\alpha}+1} \left(\frac{\partial \mathcal{S}}{\partial e_{\beta}}P_{\beta \alpha}^{\gamma}e_{\gamma} + (-1)^{\widetilde{\alpha} (\widetilde{A}+1)} \frac{\partial \mathcal{S}}{\partial x^{A}} P_{\alpha}^{A}  \right)\frac{\partial}{\partial e_{\alpha}} \in \Vect(E^{*}).
\end{eqnarray}
Thus, the vector bundle $\Pi E^{*} \rightarrow M$ is an $L_{\infty}$-algebroid. Furthermore, we see that the homological vector field is  the Hamiltonian vector field associated with the higher Schouten structure; $Q_{\mathcal{S}} = \{ \mathcal{S},\bullet \}_{P}$.
\end{enumerate}
\end{proof}

\begin{remark}
If we restrict our attention to weight two higher Poisson structures then we recover the notion of \emph{triangular Lie bialgebroids}. That is $(E,E^{*})$ is a Lie bialgebroid, see \cite{roytenberg-1999,Voronov:2001qf} for  convenient descriptions of bialgebroids.
\end{remark}

\begin{corollary}
Consider a Lie algebra $\mathfrak{g}$  that comes equipped with a higher Poisson structure $\mathcal{P} \in C^{\infty}(\Pi \mathfrak{g}^{*})$, then $\mathfrak{g}^{*}$ is canonically an $L_{\infty}$-algebra. Similarly, if $\mathfrak{g}$ is equipped with a higher Schouten $\mathcal{S} \in C^{\infty}(\mathfrak{g}^{*})$ structure, then $\Pi \mathfrak{g}^{*}$ is canonically an $L_{\infty}$-algebra.
\end{corollary}

\begin{remark}
As expected higher Poisson/Schouten structures represent a generalisation of the notion triangular Lie bialgebras \cite{Belavin1982}. The classical $R$-matrix is analogous  to the Poisson/Schouten structures and  the classical Yang--Baxter equation to the self-commutation condition or master equation.
\end{remark}

\begin{theorem}{Theorem}
Let $E \rightarrow M$ be a Lie algebroid.
\begin{enumerate}
\item If $E$ is  equipped with a higher Poisson structure $\mathcal{P} \in C^{\infty}(\Pi E^{*})$ then the algebra of Lie algebroid forms, $C^{\infty}(\Pi E)$ is a homotopy Schouten algebra.
\item If $E$ is  equipped with a higher Schouten structure $\mathcal{S} \in C^{\infty}( E^{*})$ then  the algebra of Lie algebroid forms, $C^{\infty}(\Pi E)$ is a homotopy Poisson algebra.
\end{enumerate}
\end{theorem}

\begin{proof}
\begin{enumerate}
\item The higher Schouten structure on the total space of $\Pi E$ is supplied by

\begin{eqnarray} \label{higher schouten_Poisson}
 \mathcal{S}_{\mathcal{P}} = (R^{-1})^{*}\mathcal{H}_{\mathcal{P}} &=& \left(\frac{\partial \mathcal{P}(x,\pi)}{\partial \pi_{\alpha}}S_{\alpha}^{A}  \right)p_{A}\\
\nonumber &+& (-1)^{\widetilde{\alpha}} \left((-1)^{\widetilde{A}(\widetilde{\alpha}+1)}\frac{\partial \mathcal{P}(x,\pi)}{\partial x^{A}}S_{\alpha}^{A} + \frac{\partial \mathcal{P}(x,\pi)}{\partial \pi_{\beta}}S_{\beta \alpha}^{\gamma} \pi_{\gamma}  \right)\xi^{\alpha} \\
\nonumber &\in &  C^{\infty}(T^{*}(\Pi E)),
\end{eqnarray}

\noindent where $R: T^{*}(\Pi E^{*}) \rightarrow T^{*}(\Pi E)$ is the canonical double vector bundle morphism. We have used the short hand $(R^{-1})^{*}\mathcal{P} = \mathcal{P}(x,\pi)$. Recall that the canonical double vector bundle morphism is a symplectomorphism and as such, $\mathcal{S}_{\mathcal{P}}$ defines a genuine higher Schouten structure. The higher Schouten brackets are then defined as

\begin{equation}
(\alpha_{1}, \alpha_{2}, \cdots, \alpha_{r})_{\mathcal{{P}}} = \left.\{ \cdots \{ \{ \mathcal{S}_{\mathcal{P}}, \alpha_{1} \}, \alpha_{2} \},\cdots, \alpha_{r} \}\right|_{\Pi E},
\end{equation}

\noindent with $\alpha_{I} \in C^{\infty}(\Pi E)$ and where the brackets $\{ \bullet, \bullet \}$ are the canonical Poisson brackets on $T^{*}(\Pi E)$.

\item The higher Poisson structure on the total space of $\Pi E$ is provided by

\begin{eqnarray}
 \mathcal{P}_{\mathcal{S}} = (R^{-1})^{*}\mathcal{H}_{\mathcal{S}} &=& - \left(\frac{\partial \mathcal{S}(x, \eta^{*})}{\partial \eta^{*}_{\alpha}} P_{\alpha}^{A} \right)x^{*}_{A}\\
\nonumber &+& (-1)^{\widetilde{\alpha}+1} \left( \frac{\partial \mathcal{S}(x,\eta^{*})}{\partial \eta^{*}_{\beta}} P_{\beta \alpha}^{\gamma}\eta^{*}_{\gamma} + (-1)^{\widetilde{\alpha}(\widetilde{A}+1)} \frac{\partial \mathcal{S}(x,\eta^{*})}{\partial x^{A}}P_{\alpha}^{A} \right)\eta^{\alpha} \\
\nonumber &\in& C^{\infty}(\Pi T^{*}(\Pi E)),
\end{eqnarray}

\noindent where $R : \Pi T^{*}(E^{*}) \rightarrow \Pi T^{*}(\Pi E)$ is the canonical double vector bundle morphism. We have used the shorthand $(R^{-1})^{*} \mathcal{S} = \mathcal{S}(x,\eta^{*})$. As the canonical double vector bundle morphism is an odd symplectomorphism, the structure $\mathcal{P}_{\mathcal{S}}$ is a genuine higher Poisson structure. The higher Poisson brackets are defined as

\begin{equation}
\{\alpha_{1}, \alpha_{2}, \cdots , \alpha_{r}  \}_{\mathcal{S}} = \left.   \SN{ \cdots \SN{\SN{\mathcal{P}_{\mathcal{S}}, \alpha_{1} }, \alpha_{2}},\cdots , \alpha_{r} }  \right|_{\Pi E},
\end{equation}

\noindent with $\alpha_{I} \in C^{\infty}(\Pi E)$ and where the brackets $\SN{ \bullet, \bullet }$ are the canonical Schouten brackets on $\Pi T^{*}(\Pi E)$.
\end{enumerate}
\end{proof}

\noindent The series of brackets on $C^{\infty}(\Pi E)$ should be thought of as simultaneously  the Lie algebroid and homotopy generalisation of the Koszul--Schouten bracket in classical Poisson geometry \cite{Koszul;1985}. (The earliest construction of a bracket on one-forms over a symplectic manifold can be found in  the first edition of Abraham \& Marsden \cite{Abraham:1994}). The example of higher Poisson structures has already been discussed by Khudaverdian \& Voronov \cite{khudaverdian-2008}.\\

 \begin{corollary} If the  smooth functions over a manifold form  a homotopy Poisson/Schouten algebra, then the space of differential forms over the manifold  is canonically  a homotopy Schouten/Poisson algebra.
 \end{corollary}

\subsection{Higher Koszul--Schouten brackets on Lie algebroids}\label{higher koszul--schouten}

\noindent In this subsection we define another series of brackets on $C^{\infty}(\Pi E)$ when the Lie algebroid $E \rightarrow M$ is equipped with a higher Poisson structure. We do this by following Koszul's original construction \cite{Koszul;1985} and define the series of brackets as higher brackets generated by an algebroid analogue of the  Koszul--Brylinski operator \cite{Brylinski1988,Koszul;1985}. This will provide a homotopy BV-algebra structure on the space of Lie algebroid forms. The associated series of higher antibrackets will be known as higher Koszul--Schouten brackets as to distinguishes them from the earlier Schouten brackets.\\

\noindent Before we do this, we recall the basic elements of the Cartan calculus on Lie algebroids \cite{Kosmann-Schwarzbach-1990}. The Cartan calculus consists of three operators in $\End(C^{\infty}(\Pi E))$: The de Rham differential, interior product and Lie derivative that act on Lie algebroid forms viz

\begin{enumerate}
\item The de Rham differential \footnote{we change notation slightly to reflect the relation with the standard Cartan calculus.}
\begin{equation}
d_{E} = Q =  \xi^{\alpha}Q_{\alpha}^{A}\frac{\partial }{\partial x^{A}} + \frac{1}{2} \xi^{\alpha}\xi^{\beta}Q_{\beta \alpha}^{\gamma} \frac{\partial}{\partial \xi^{\gamma}}.
\end{equation}
Note $\widetilde{d_{E}} = 1$.
\item The interior product
\begin{equation}
i_{X} = (-1)^{\widetilde{X}} \sum_{k = 0}^{\infty} \frac{1}{k!} X^{\alpha_{1} \cdots \alpha_{k}} \frac{\partial}{\partial \xi^{\alpha_{k}}}\cdots \frac{\partial}{\partial \xi^{\alpha_{1}}},
\end{equation}
which can be understood as the assignment of a differential operator $X\rightsquigarrow i_{X}$, given a Lie algebroid multivector field $X = \sum_{k=0}^{\infty} \frac{1}{k!} X^{\alpha_{1} \cdots \alpha_{k}}(x) \eta_{\alpha_{k}}\cdots \eta_{\alpha_{1}} \in C^{\infty}(\Pi E^{*})$. We have picked natural coordinates $\{x^{A}, \eta_{\alpha} \}$ on $\Pi E^{*}$. The infinite sum is understood formally. Note this assignment is even: $\widetilde{i_{X}} = \widetilde{X}$.
\item The Lie derivative
\begin{eqnarray}\label{lie derivative}
 L_{X} &=& [d_{E}, i_{X}]\\
\nonumber  &=& \sum_{k = 0}^{\infty}\left( \left( (-1)^{\widetilde{X}}\frac{1}{k!} \xi^{\alpha}Q_{\alpha}^{A} \frac{\partial X^{\sigma_{1} \cdots \sigma_{k}}}{\partial x^{A}}\right.\right.\\
\nonumber & -& \left. \left. (-1)^{(\widetilde{\gamma}+ \widetilde{\sigma_{1}}+1)(\widetilde{X}+ \widetilde{\gamma}+1)} \frac{1}{(k-1)!}\xi^{\beta}Q_{\beta \gamma}^{\sigma_{1}} X^{\gamma \sigma_{2} \cdots \sigma_{k}} \right) \frac{\partial}{\partial \xi^{\sigma_{k}}} \cdots\frac{\partial}{\partial \xi^{\sigma_{1}}}   \right.\\
\nonumber &-& \left. (-1)^{(\widetilde{A}+ \widetilde{\sigma_{1}})(\widetilde{X}+ \widetilde{\sigma_{1}}+1)} \frac{1}{(k-1)!}Q_{\sigma_{1}}^{A}X^{\sigma_{1} \cdots \sigma_{k}} \frac{\partial}{\partial \xi^{\sigma_{k}}}\cdots \frac{\partial }{\partial \xi^{\sigma_{2}}} \frac{\partial}{\partial x^{A}} \right).
\end{eqnarray}
Note that this assignment $X \rightsquigarrow L_{X}$ is an odd:  $\widetilde{L_{X}} = \widetilde{X}+1$.
\end{enumerate}

\begin{remark}
On the antitangent bundle $\Pi TM$ of a manifold $M$, the above reduced to the generalised Cartan calculus involving multivector fields \cite{Tulczjyew:1974B} up on a shift in parity.
\end{remark}
\noindent The respective weights (relative to the natural weight on $\Pi E^{*}$) are $\w(d_{E})= -1$, $\w(i_{X})= \w(X)$ and $\w(L_{X})= \w(X)-1$, assuming the Lie algebroid multivector field is homogenous in weight. Further note that if the Lie algebroid multivector field in question is homogenous in weight, say $\w(X) =r$ then the interior product and Lie derivative are both differential operators of order $r$. Crucially, they do not satisfy the Leibnitz identity apart from the isolated case of $r=1$. \\

\noindent The above endomorphisms satisfy a series of identities which we will refer to as the Cartan identities:

\begin{subequations}
\begin{eqnarray}
 d_{E}^{2}&=& 0,\\
 \left[d_{E}, L_{X} \right] &=& 0,\\
 \left[i_{X}, i_{Y} \right] &=& 0,\\
 i_{\SN{X,Y}_{S}} &=&\left[ i_{X}, L_{Y} \right],\\
 L_{\SN{X,Y}_{S}} &=& \left[L_{X}, L_{Y}  \right],\\
L_{YX} &=& L_{Y} \circ i_{X} + (-1)^{\widetilde{Y}}i_{Y} \circ L_{X},
\end{eqnarray}
\end{subequations}

\noindent All the above can be proved by direct calculation via local coordinates, \cite{Kosmann-Schwarzbach-1990,Tulczjyew:1974B}.

\begin{definition}{Definition}
Let $E \rightarrow M$ be a Lie algebroid with higher Poisson structure \newline $\mathcal{P} \in C^{\infty}(\Pi E^{*})$. The Koszul--Brylinski operator  is defined as the Lie derivative along the higher Poisson structure:  $\Delta_{\mathcal{P}} := L_{\mathcal{P}} = d_{E}\circ i_{\mathcal{P}} - i_{\mathcal{P}} \circ d_{E}$.
\end{definition}

\noindent That is we have the association of a differential operator acting on Lie algebroid forms given a higher Poisson structure $\mathcal{P} \rightsquigarrow L_{\mathcal{P}}$ viz Eqn. \ref{lie derivative}.\\

\begin{theorem}{Theorem}
Let $E \rightarrow M$ be a Lie algebroid equipped with a higher Poisson structure $\mathcal{P} \in C^{\infty}(\Pi E^{*})$. The pair $(C^{\infty}(\Pi E), \Delta_{\mathcal{P}})$ is a homotopy BV-algebra.
\end{theorem}

\begin{proof}
 Recall Definition \ref{Def homotopy BV algebra}.  The fact that the   Koszul--Brylinski  operator is odd follows from the fact that the higher Poisson structure is even. Using the Cartan identities it is clear that $(L_{\mathcal{P}})^{2} = \frac{1}{2}[L_{\mathcal{P}}, L_{\mathcal{P}}] = \frac{1}{2}L_{\SN{\mathcal{P}, \mathcal{P} }_{S}}=0 \Leftrightarrow \SN{\mathcal{P}, \mathcal{P}}_{S}=0$. Thus the operator $\Delta_{\mathcal{P}}$ ``squares to zero".
\end{proof}

\begin{definition}{Definition}
Consider a Lie  algebroid $E \rightarrow M$ with a higher Poisson structure $\mathcal{P}\in C^{\infty}(\Pi E^{*})$. The associated higher Koszul--Schouten brackets between Lie algebroid forms are defined as the brackets generated by the Koszul--Brylinski operator
\begin{equation}
[\alpha_{1}, \alpha_{2}, \cdots , \alpha_{r}]_{\mathcal{P}} := \left[\cdots\left[ \left[\Delta_{\mathcal{P}}, \alpha_{1}  \right], \alpha_{2} \right], \cdots, \alpha_{r} \right]\mathds{1},
\end{equation}
with $\alpha_{I} \in C^{\infty}(\Pi E)$ and $\mathds{1}$ is the unit Lie algebroid form, i.e. the function on $M$ of constant value one.
\end{definition}

\begin{warning}
This series of odd brackets are not higher Schouten brackets as they do not satisfy a Leibnitz rule over the supercommutative product of Lie algebroid forms. Instead we have the well-known recursive relation \cite{Akman:1995tm,Koszul;1985}
\begin{eqnarray}
 [\alpha_{1}, \cdots, \alpha_{r-1}, \alpha_{r}\alpha_{r+1}]_{\mathcal{P}} &=& [\alpha_{1}, \cdots, \alpha_{r-1}, \alpha_{r}]_{\mathcal{P}}\alpha_{r+1}\\
\nonumber  &+&(-1)^{(\widetilde{\alpha}_{1}+ \cdots + \widetilde{\alpha}_{r-1}+1)\widetilde{\alpha}_{r}}\alpha_{r}[\alpha_{1}, \cdots, \alpha_{r+1}]_{\mathcal{P}}\\
\nonumber &+& [\alpha_{1}, \cdots, \alpha_{r-1}, \alpha_{r}, \alpha_{r+1}]_{\mathcal{P}}.
\end{eqnarray}
\end{warning}

\begin{remark}
There is no direct analogue for higher Schouten structures and the higher Poisson brackets on Lie algebroid forms. Recall that the series of higher Poisson brackets have even/odd parity for an even/odd number of arguments. It is known that no operator can generate such a series of brackets that satisfy the Jacobiators needed to form an $L_{\infty}$-algebra.
\end{remark}

\noindent The higher Schouten brackets presented in Section (\ref{higher Poisson and Schouten}) and higher Koszul--Schouten brackets presented here are not independent constructions. The higher Schouten brackets can be viewed as a ``classical limit" of the higher Koszul--Schouten brackets (the original idea is due to Voronov \cite{voronov-2004}). To see this we first define a \emph{deformed} higher Poisson structure as

\begin{equation}
\mathcal{P} \rightsquigarrow \mathcal{P}[\hbar] = \sum_{k=0}^{\infty}\frac{(\hbar)^{k}}{k!}\mathcal{P}^{\alpha_{1} \cdots \alpha_{k}}\eta_{\alpha_{k}}\cdots \eta_{\alpha_{1}}.
\end{equation}

\noindent Here we think of $\hbar$ as a formal even weight zero deformation parameter. Specifically, no physical significance (eg. reference to some quantisation procedure) is attached to this parameter.

\begin{theorem}{Theorem}
The higher Schouten brackets are given by a formal classical limit of the higher Koszul--Schouten brackets generated by the deformed  Koszul--Brylinski operator. That is
\begin{eqnarray}
\nonumber (\alpha_{1}, \alpha_{2}, \cdots ,\alpha_{r})_{\mathcal{P}} &=& \lim_{\hbar \rightarrow 0  } \hbar^{-r} [\alpha_{1}, \alpha_{2},\cdots , \alpha_{r}]_{\mathcal{P}[\hbar]}\\
&=&  \lim_{\hbar \rightarrow 0} \hbar^{-r} \left[\cdots\left[ \left[\Delta_{\mathcal{P[\hbar]}}, \alpha_{1}  \right], \alpha_{2} \right], \cdots, \alpha_{r} \right]\mathds{1}.
\end{eqnarray}
\end{theorem}

\begin{proof}
Consider the map $\sigma L_{X} = L_{X}(x, \eta, p, \pi) \in C^{\infty}(T^{*}(\Pi E))$ defined by $\frac{\partial}{\partial x^{A}} \rightarrow p_{A}$ and $\frac{\partial}{\partial \xi^{\alpha}} \rightarrow \pi_{\alpha}$ for an arbitrary $X \in C^{\infty}(\Pi E^{*})$. As the components of the Lie derivative are tensorial under double vector bundle morphisms this map is well-defined. It is the total symbol of the Lie derivative (see for example \cite{Hormander:1985III}).  The total symbol for the higher Koszul--Brylinski operator  is given by $ \sigma \Delta_{\mathcal{P}} = \mathcal{S}_{\mathcal{P}}$, see Eqn.(\ref{higher schouten_Poisson}). Then it is clear that
\begin{equation}
\sigma [\Delta_{\mathcal{P}}, \Delta_{\mathcal{P}}] = \{\mathcal{S}_{\mathcal{P}},\mathcal{S}_{\mathcal{P}} \}=0,
\end{equation}
as taking the total symbol  takes commutators to Poisson brackets on $T^{*}(\Pi E)$. As the only non-trivial contribution to the deformed Koszul--Schouten brackets are from the weight greater  than $r$ components of $\mathcal{P}[\hbar]$ we see that
\begin{equation}
\nonumber \lim_{\hbar \rightarrow 0} \left(\frac{1}{\hbar}\right)^{r} \left[\cdots\left[ \left[\Delta_{\mathcal{P[\hbar]}}, \alpha_{1}  \right], \alpha_{2} \right], \cdots, \alpha_{r} \right]\mathds{1} = \left[\cdots\left[ \left[L_{\stackrel{r}{\mathcal{P}}}, \alpha_{1}  \right], \alpha_{2} \right], \cdots, \alpha_{r} \right]\mathds{1},
\end{equation}
where $\stackrel{r}{\mathcal{P}} = \frac{1}{r!}\mathcal{P}^{\alpha_{1} \cdots \alpha_{r}}\eta_{\alpha_{r}} \cdots \eta_{\alpha_{1}}$ is the $r$-th component of the higher Poisson structure.  Taking the total symbol gives
\begin{eqnarray}
\sigma \left(\left[\cdots\left[ \left[L_{\stackrel{r}{\mathcal{P}}}, \alpha_{1}  \right], \alpha_{2} \right], \cdots, \alpha_{r} \right]\mathds{1}\right) &=& \left\{\cdots\left\{ \left\{\sigma L_{\stackrel{r}{\mathcal{P}}}, \alpha_{1}  \right\}, \alpha_{2} \right\}, \cdots, \alpha_{r} \right\}\\
\nonumber &=& \left. \left\{\cdots\left\{ \left\{\mathcal{S}_{\mathcal{P}}, \alpha_{1}  \right\}, \alpha_{2} \right\}, \cdots, \alpha_{r} \right\}\right|_{\Pi E \subset T^{*}(\Pi E)}.
\end{eqnarray}
 Remember that each component of Lie derivative is invariant under (graded) double vector bundle morphisms.  Putting this together we see that
\begin{equation}
\lim_{\hbar \rightarrow 0  } \left(\frac{1}{\hbar}\right)^{r}[\alpha_{1}, \alpha_{2},\cdots , \alpha_{r}]_{\mathcal{P}[\hbar]} = \left. \left\{\cdots\left\{ \left\{ \mathcal{S}_{\mathcal{P}}, \alpha_{1}  \right\}, \alpha_{2} \right\}, \cdots, \alpha_{r} \right\}\right|_{\Pi E \subset T^{*}(\Pi E)},
\end{equation}
and the result is established.
\end{proof}

\noindent Via these constructions we interpret the higher Schouten structure $\mathcal{S}_{\mathcal{P}} \in C^{\infty}(T^{*}(\Pi E))$ as a ``Hamiltonian function" associated with the  Koszul--Brylinski operator. Thus, both the higher Schouten and higher Koszul--Schouten brackets on Lie algebroid forms can be traced back to the Lie derivative along the higher Poisson structure.\\

\begin{corollary}
If the smooth functions over a manifold form a homotopy Poisson algebra, then the space of differential forms over the manifold is canonically a homotopy BV-algebra. Furthermore, the  higher Schouten brackets between differential forms are the ``classical limit" of the higher Koszul--Schouten brackets.
\end{corollary}


\section{Discussion}\label{discussion}
\noindent Lie algebroids have by now become an established part of modern geometry and mathematical physics.  From the point of view of this work, the main interest in Lie algebroids lies in the fact they present a unification of Schouten \& Poisson structures as well as a natural  setting to discuss (higher) brackets on supermanifolds.  This may in turn be of interest in theoretical physics, in particular things related to the Batalin--Vilkovisky formalism \cite{Batalin:1981jr,Batalin:1984jr}. \\

\noindent Indeed the initial motivation for this work lies in odd symplectic geometry, and in particular the work of Khudaverdian \& Voronov \cite{khudaverdian-2006}. By thinking of  symplectic structures as functions on $\Pi TM$ and the associated Poisson or Schouten structures as functions on $\Pi T^{*}M$ or $T^{*}M$ respectively, one is lead  (ignoring the weight) to graded analogues of the classical Tulczyjew triple for $M = T^{*}N$ and $M = \Pi T^{*}N$.  This allows for aspects of odd symplectic geometry to be included in Tulczyjew's constructions.\\

\noindent Tulczyjew's motivation was to geometrically understand aspects of classical mechanics and in particular the Legendre transform \cite{Tulczyjew:1974}.   However, the question of what aspects of the BV-antifield formalism can be restated in terms of the graded triples (or some variant of) presented in this paper remains unexplored.\\

\noindent Furthermore, the physical application of the various higher brackets presented in this paper remains somewhat elusive. If nothing else, they represent clear geometric examples of $L_{\infty}$-algebras. However, the structure of a higher Poisson structure can be found in the classical BV-antifield formulism (mod extra gradings). It is quite possible that the associated higher Schouten and higher Koszul--Schouten brackets have some physical interpretation in terms of the BV-antifield formalism.  \\

\noindent The ``graded super" constructions presented in this work give a nice setting to discuss  Poisson and Schouten structures on Lie algebroids. It must be remarked that in the context of classical Poisson geometry, Cattaneo \& Zambon \cite{Cattaneo2009} have used graded geometry to investigate Poisson reduction. Presumably, many other geometric structures, such as twisted Poisson, Poisson--Nijenhuis, Jacobi and similar on Lie algebroids can be restated in this graded context. Works on various geometric structures in terms of graded manifolds  include \cite{Antunes2008,Grabowski2009,Kosmann-Schwarzbach2007,Mehta2006,Roytenberg:2001}  as well as many others.

\section*{Acknowledgments}

\noindent The author would like to thank  Th.Th. Voronov and H.M. Khudaverdian for many interesting discussions and guidance. An honorable mention must go to  J. Grabowski and P. Urba$\textnormal{\'n}$ski for help with bibliographical issues.  A special thank you goes to  K.C.H. Mackenzie. The author also like to thank the two anonymous referees whose comments served to improve the presentation of this work. This work was funded by the British Tax-payer via an EPSRC DTA.

\bibliography{preprintbib}

\end{document}